\documentclass{aa}  

\usepackage{graphicx}
\usepackage{txfonts}
\graphicspath{{./}{figures/}}

\newcommand{\numax}{\mbox{$\nu_{\rm max}$}}

\newcommand{\kepler}[0]{\emph{Kepler}}
\newcommand{\corot}[0]{CoRoT}

\newcommand{\echelle}{{\'e}chelle}
\newcommand{\Dnu}{\mbox{$\Delta\nu$}}
\newcommand{\baql}{\mbox{$\beta$~Aql}}
\newcommand{\muHz}{\mbox{$\mu$Hz}}
\newcommand{\ms}{\mbox{ms$^{-1}$}}
\newcommand{\Teff}{\mbox{$T_\text{eff}$}}
\newcommand{\Rsolar}{\mbox{$\text{R}_{\odot}$}}
\newcommand{\Msolar}{\mbox{$\text{M}_{\odot}$}}

\usepackage{xcolor}

\newcommand{\new}[1]{{\color{red}\textbf{#1}}}

\newif\ifarxiv
\arxivfalse
\arxivtrue 

\ifarxiv
    \renewcommand{\new}[1]{#1} 
\fi
\begin{document}

\title{Asteroseismology of the G8 subgiant $\beta$ Aquilae with SONG-Tenerife, SONG-Australia and TESS}


\author{
Hans Kjeldsen\inst{1} \and
Timothy R. Bedding\inst{2} \and
Yaguang Li\inst{3} \and
Frank Grundahl\inst{1} \and
Mads Fredslund Andersen\inst{1} \and
Duncan J. Wright\inst{4} \and
Jack Soutter\inst{4} \and
Robert Wittenmyer\inst{4} \and 
Claudia Reyes\inst{5} \and
Dennis Stello\inst{5} \and
Courtney Crawford\inst{2} \and
Yixiao Zhou\inst{6} \and \\
Mathieu Clerte\inst{4} \and 
Pere L. Pall{\'e}\inst{7,8} \and
Sergio Simon-Diaz\inst{7,8} \and
J{\o}rgen Christensen-Dalsgaard\inst{1} \and
Rasmus Handberg\inst{1} \and \\
Hasse Hansen\inst{1} \and
Paul Heeren\inst{1,9} \and 
Jens Jessen-Hansen\inst{1} \and
Mikkel~N.~Lund\inst{1} \and
Mia S. Lundkvist\inst{1} \and
Karsten Brogaard\inst{1} \and \\
Ren{\'e} Tronsgaard\inst{1} \and 
Jonatan Rudrasingam\inst{2} \and 
Luca Casagrande\inst{10} \and
Jonathan Horner\inst{3} \and 
Daniel Huber\inst{2,3} \and
John Lattanzio\inst{11} \and 
Sarah L. Martell\inst{5} \and
Simon J. Murphy\inst{4} 
}

\institute{
Stellar Astrophysics Centre, Department of Physics and Astronomy, Aarhus University, DK-8000 Aarhus C, Denmark
\email{hans@phys.au.dk}
\and 
Sydney Institute for Astronomy, School of Physics, University of Sydney, NSW 2006, Australia
\and 
Institute for Astronomy, University of Hawai‘i, 2680 Woodlawn Drive, Honolulu, HI 96822, USA
\and 
Centre for Astrophysics, University of Southern Queensland, Toowoomba, QLD 4350, Australia
\and 
School of Physics, University of New South Wales, NSW 2052, Australia
\and 
School of Physical and Chemical Sciences, Te Kura Mat{\=u}, University of Canterbury, Christchurch 8140, Aotearoa, New Zealand
\and 
Instituto de Astrof{\'i}sica de Canarias, 38200 La Laguna, Tenerife, Spain
\and 
Departamento de Astrof{\'i}sica, Universidad de La Laguna (ULL), 38206 La Laguna, Tenerife, Spain
\and 
Landessternwarte, Zentrum f{\"u}r Astronomie der Universität Heidelberg,
K{\"o}nigstuhl 12, 69117 Heidelberg, Germany
\and 
Research School of Astronomy and Astrophysics, Australian National University, Canberra, ACT 2611, Australia
\and 
School of Physics and Astronomy, Monash University, Clayton, Australia
}

  \abstract
   {}
   {We present time-series radial velocities of the G8 subgiant star \baql\ obtained in 2022 and 2023 using SONG-Tenerife and\new{, for the first time,} SONG-Australia. We also analyse a sector of TESS photometry that overlapped with the 2022 SONG data.}
   {We processed the time series to assign weights and to remove bad data points. The resulting power spectrum clearly shows solar-like oscillations centred at 430\,\muHz. The TESS light curve shows the oscillations at lower signal-to-noise, reflecting the fact that photometric measurements are much more affected by the granulation background than are radial velocities.}
   {The simultaneous observations in velocity and photometry represent the best such measurements for any star apart from the Sun. They allowed us to measure the ratio between the bolometric photometric amplitude and the velocity amplitude to be $26.6 \pm 3.1$\,ppm/\ms. We measured this ratio for the Sun from published SOHO data to be $19.5 \pm 0.7$\,ppm/\ms\ and, after accounting for the difference in effective temperatures of \baql\ and the Sun, these values align with expectations.  In both the Sun and \baql, the photometry-to-velocity ratio appears to be a function of frequency.  We also measured the phase shift of the oscillations in \baql\ between SONG and TESS to be $-113^{\circ} \pm 7^{\circ}$, which agrees with the value for the Sun and also with a 3-D simulation of a star with similar properties to \baql.  Importantly for exoplanet searches, we argue that simultaneous photometry can be used to predict the contribution of oscillations to radial velocities.  We measured frequencies for \new{22} oscillation modes in \baql\ and carried out asteroseismic modelling, yielding an excellent fit to the frequencies. We derived accurate values for the mass and age, and were able to place quite strong constraints on the mixing-length parameter. Finally, we show that the oscillation properties of \baql\ are very similar to stars in the open cluster M67.
   }
   {}

   \keywords{Asteroseismology -- 
   Stars: individual: $\beta$ Aql -- 
   Stars: late-type -- 
   Stars: oscillations (including pulsations) -- 
   Techniques: radial velocities}

   \maketitle
%

\section{Introduction}

Before the era of space photometry, most detections of solar-like oscillations were made with radial-velocity measurements \citep[see reviews by][]{Bedding+Kjeldsen2003, Elsworth+Thompson2004, Cunha++2007, Aerts++2008, Aerts++2010-book, Bedding2014}.
Starting in 2009, asteroseismology of solar-like oscillations underwent a revolution thanks to the photometric space missions \corot, \kepler\ and TESS \citep[see reviews by][]{Chaplin+Miglio2013, Garcia+Ballot2019, Jackiewicz2021}.
These space missions have provided long and nearly uninterrupted light curves with exquisite photometric precision for thousands of stars.

Importantly, measurements of solar-like oscillations with space-based photometry are usually limited by the intrinsic noise from the stars themselves, which arises from photometric fluctuations caused by surface granulation \citep[e.g.,][]{Chaplin++2011-detectability, Karoff++2013, Samadi++2013, Kallinger++2014, Campante++2016, Pande++2018, RodriguezDiaz++2022}.
Radial velocity (RV) measurements are much less affected because the granulation background is much weaker relative to the oscillations than in photometry \citep{Harvey1988, Grundahl++2007, Kjeldsen+Bedding2011, Luhn++2025}. Indeed, recent observations with the VLT and Keck have shown that RV observations can measure the low-amplitude oscillations of K dwarfs in only a few nights \citep{Campante++2024, Hon++2024, Lundkvist++2024, Li-Yaguang++2025}. On the other hand, RV observations with ground-based telescopes will inevitably have gaps in the time series that compromise oscillation measurements, and the benefits of having two or more telescopes are well-established \citep[e.g.,][]{Arentoft++2014}.

The primary aim of SONG (Stellar Observations Network Group) is to collect high-precision RV measurements of oscillating stars. The first node at Observatorio del Teide on Tenerife, Spain, has been operating since 2014 and consists of the 1-m Hertzsprung SONG Telescope and a coud{\'e} \echelle\ spectrograph with an iodine cell \citep{Grundahl++2007, Andersen++2014, Andersen++2016}. In this paper, we present SONG results that include a new node in Australia, at Mount Kent Observatory near Toowoomba in Queensland.
SONG-Australia comprises two 0.7-m telescopes that are identical to those used by the MINERVA-Australis facility on the same site \citep{Addison++2019}.  Fibres from the two SONG telescopes are fed to a high-resolution spectrograph, equipped with an iodine cell, that is similar to the one at SONG-Tenerife.  

The target for these observations was the bright G8 subgiant $\beta$~Aquilae. This star is particularly interesting because it is in transition from a subgiant to a red giant.    Our analysis combines RV observations from the SONG nodes in Tenerife and Australia with photometry from TESS.

\section{Properties of \baql}
\label{sec:properties}

The G8 subgiant \baql\ (HR 7602; HD 188512; HIP 98036; TIC 375621179) is a Gaia benchmark star \citep{Soubiran++2024}.
Its angular diameter has been measured by interferometry using both VLTI/PIONIER \citep{Rains++2020} and CHARA/PAVO \citep{Karovicova++2022}, giving angular diameters (corrected for limb darkening) of $2.133 \pm 0.012$\,mas and $2.096 \pm 0.014$\,mas, respectively, and the effective temperature as $5071 \pm 37$\,K and $5113 \pm 20$\,K.  For this work we adopt the mean of these two temperatures with a conservative error bar: $\Teff = 5092 \pm 50$\,K. This agrees with the spectroscopic value of $\Teff = 5030 \pm 80$\,K measured by \citet{Bruntt++2010}. Those authors reported a metallicity for \baql\ of $\text{[Fe/H]} = -0.21 \pm 0.07$, which we adopt for the modelling described in Sec.~\ref{sec:models}.

\baql\ is in a wide binary with an M-dwarf companion (\baql~B) that is separated by 13.4\,arcsec \citep{Mason++2001, Karovicova++2022}, with a projected physical separation of 180.5\,au \citep{El-Badry++2021}.  The Gaia magnitudes of the two components are $G=3.48$ and $G=10.47$, which means the M dwarf makes a negligible contribution to the flux. Importantly, the Gaia DR3 parallax of the secondary ($73.3889 \pm 0.0215$\,mas) is much more precise than that of the primary ($73.52 \pm 0.14$\,mas; \citealt{Gaia2021}). This is presumably because the primary is so bright, and leads us to prefer the parallax of the secondary as a measure of the distance to the system. Combining the Gaia DR3 distance to \baql~B of $13.617 \pm 0.004$\,pc \citep{Bailer-Jones++2021} with the mean angular diameter of \baql~A ($2.115 \pm 0.010$\,mas; see above) gives a stellar radius of $3.096 \pm 0.015$\,\Rsolar.

\label{sec:activity}

A possible magnetic activity cycle in \baql\ was reported by \citet{Baliunas++1995}, based on variations in the calcium H\&K S-index as measured by the Mount Wilson HK-Project.  They analysed measurements from 1978 to 1991 and suggested a possible period of $4.1\pm0.1$\,yr but with a grade of `poor' based on the false-alarm probability (FAP).
Subsequently, \citet{Butkovskaya++2017} suggested an activity cycle with a period of 2.7\,yr based on a spectropolarimetric measurements.  More recently, \citet{Lee++2024} monitored variability of \baql\ in H$\alpha$ over four years (2019--2022) and found a linear trend from activity but no short-term variability from rotation.  

\begin{figure}
    \centering
    \includegraphics[width=\hsize]{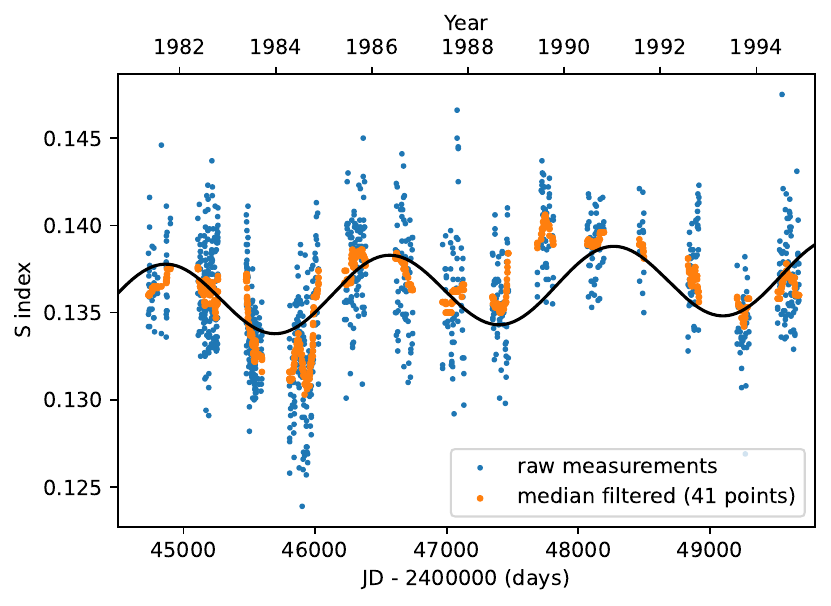}
    \caption{S-index measurements of \baql\ from the Mount Wilson HK-Project (blue circles) and the best-fitting sinusoid with a period of 4.7\,yr and a linear slope (black curve). The orange points have been smoothed with a median filter that was 41 points wide. }
    \label{fig:S-index}
    \end{figure}

The data from the Mount Wilson HK-Project used by \citet{Baliunas++1995} are now available to the research community.\footnote{\url{https://dataverse.harvard.edu/dataverse/mwo_hk_project}} 
We have downloaded the S-index measurements for \baql, which now cover 13.5\,yr from May 1981 to November 1994 (including four years subsequent to those considered by \citealt{Baliunas++1995}). The measurements are shown in Fig.~\ref{fig:S-index} and we see good evidence for a period of $4.7 \pm 0.4$ \,yr, which could reflect a magnetic activity cycle.

\section{Observations}
\label{sec:observations}

The observations we have used to perform asteroseismology on \baql\ are described in the following sections. In addition to the data from SONG (Sec.~\ref{sec:song}) and TESS (Sec.~\ref{sec:tess}), we have also analysed the RV measurements from SARG published by \citet[][see Sec.~\ref{sec:sarg}]{Corsaro++2012} and unpublished   archival RV measurements from HARPS (Sec.~\ref{sec:harps}).

\subsection{SONG (2022 and 2023)}
\label{sec:song}

\begin{figure}
    \centering
    \includegraphics[width=\hsize]{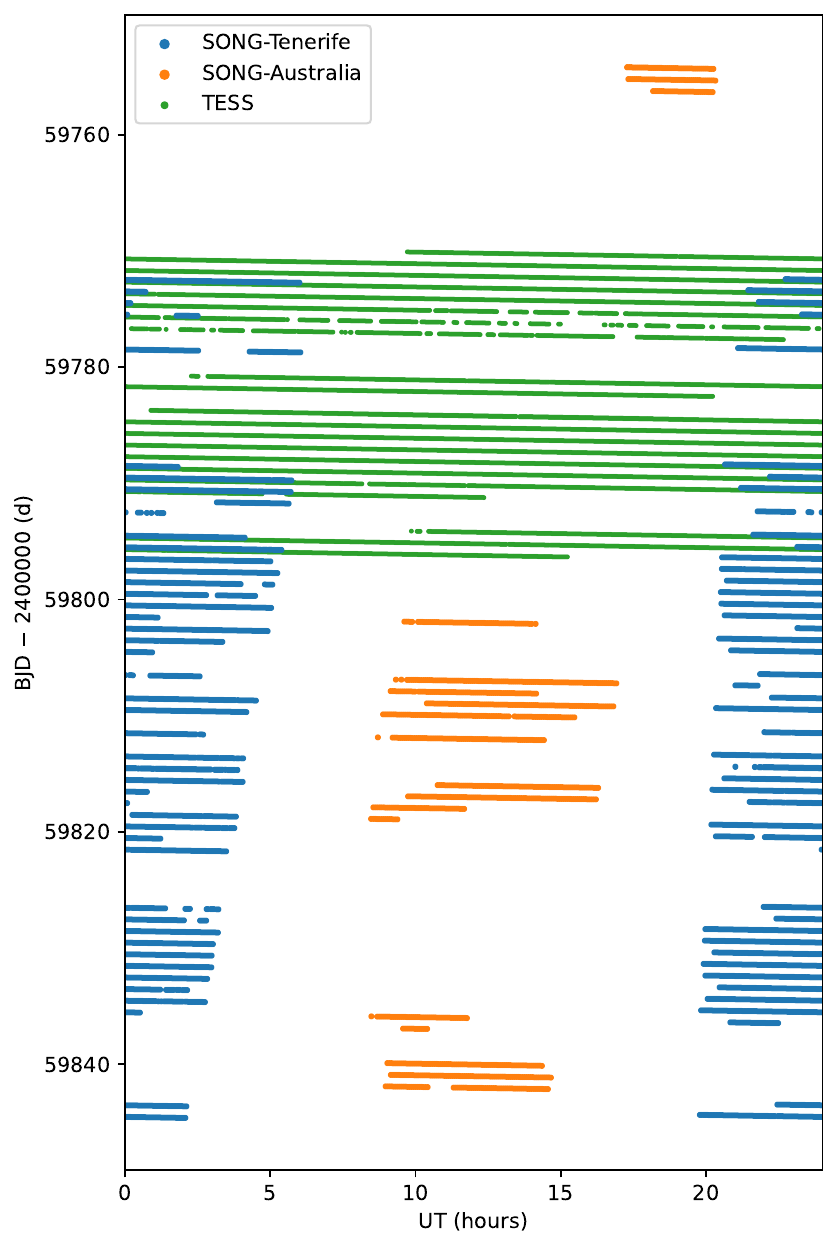}
    \caption{Daily coverage of \baql\ during the 2022 season from SONG and TESS.
    }
    \label{fig:time-series-2022-nightly}
    \end{figure}

In 2022 we observed \baql\ with SONG-Tenerife over a period spanning 74 nights (11 July to 20 September 2022) with a median sampling cadence of 185\,s.  Observations were obtained on about two-thirds of these nights, with the coverage shown by the blue points in Fig.~\ref{fig:time-series-2022-nightly}.  We obtained data with SONG-Australia on 18 nights spread over 90 nights (23 June to 20 September 2022) during a period of very poor weather with a median sampling cadence of 250\,s. These are shown by the orange points in Fig.~\ref{fig:time-series-2022-nightly}.  In 2023 we observed \baql\ with SONG-Tenerife on 33 nights spread over 58 nights (20 June to 16 August 2023) with a median sampling cadence of 126\,s. 

\begin{figure*}
    \centering
    \includegraphics[width=\hsize]{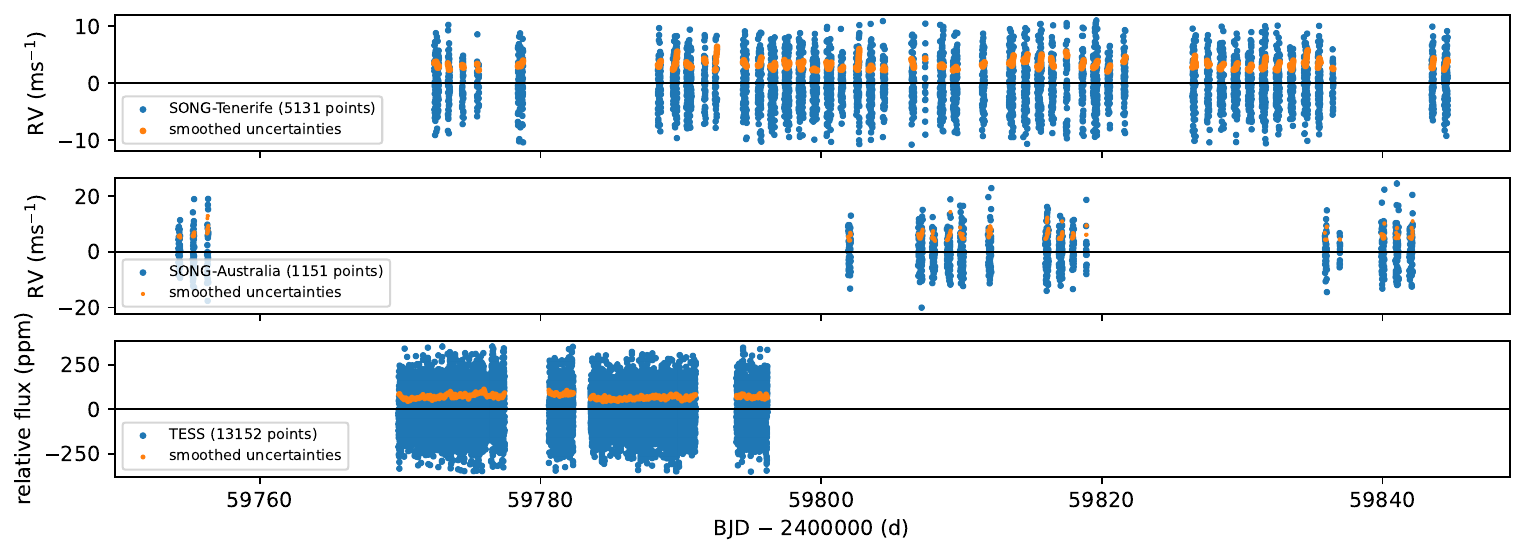}
    \caption{Time series of \baql\ from the 2022 season, showing radial velocities from SONG nodes in Tenerife (median uncertainty 3.0\,\ms) and Australia (median uncertainty 5.2\,\ms), and photometry from TESS (Sector 54; median uncertainty 68\,ppm). Processing of the velocities and their uncertainties is described in Sec.~\ref{sec:data-processing}.}
    \label{fig:time-series-2022}
    \end{figure*}

\begin{figure}
    \centering
    \includegraphics[width=\hsize]{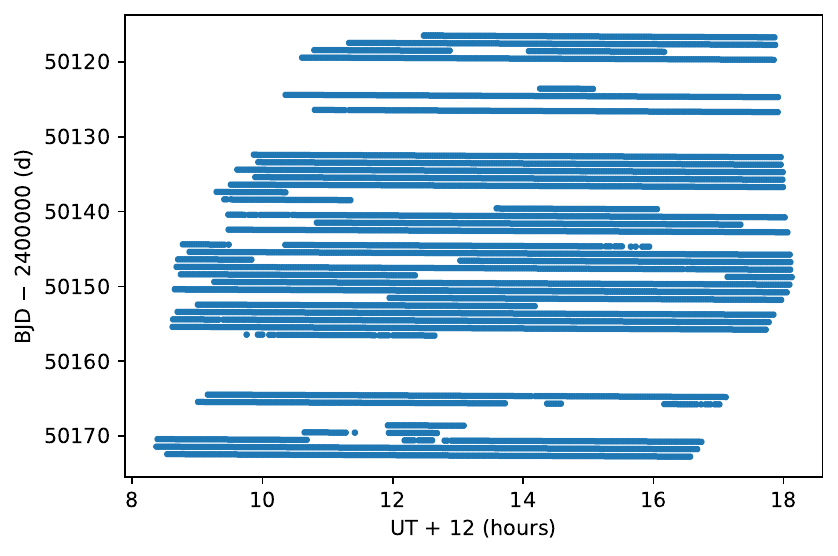}\\
    \includegraphics[width=\hsize]{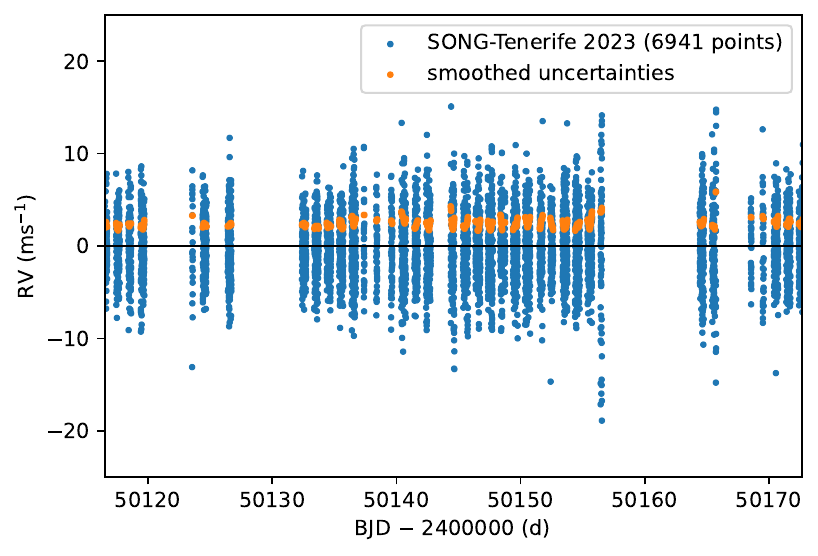}
    \caption{Time series of \baql\ from SONG Tenerife during the 2023 season. Top: nightly coverage of the measurements. Bottom: radial velocities (median uncertainty 2.3\,\ms; see Secs.~\ref{sec:song} and~\ref{sec:data-processing}).
    }
    \label{fig:time-series-2023}
    \end{figure}

Radial velocities for both SONG nodes were extracted from the spectra using the {\tt pyodine} software described by \citet{Heeren++2023}. The 2022 time series are shown in the top two panels of Fig.~\ref{fig:time-series-2022} and
the 2023 time series from SONG-Tenerife is shown in Fig.~\ref{fig:time-series-2023}.
The processing of the time series is described in Sec.~\ref{sec:data-processing}. 

\subsection{TESS (2022)}
\label{sec:tess}

The Transiting Exoplanet Survey Satellite (TESS; \citealt{Ricker++2015}) observed \baql\ in Sector~54 (8 July to 5 August 2022). The light curve had a sampling cadence of 120\,s and the temporal coverage is shown in Fig.~\ref{fig:time-series-2022-nightly} (green points).  Given the brightness of the star, we extracted the light curve ourselves from the target pixel files using the {\tt lightkurve} package \citep{lightkurve2018}.
There are 17,900 measurements spanning 26.2\,d and the processed time series is shown in the bottom panel of Fig.~\ref{fig:time-series-2022}.

We note that \citet{Corsaro++2024} analysed the TESS light curve of \baql\ as part of a larger study of magnetically active solar-like oscillators. Using the standard PDCSAP light curve,\footnote{Pre-search Data Conditioning Simple Aperture Photometry, available from the Mikulski Archive for Space Telescopes, \url{https://archive.stsci.edu/}.}
 they reported $\numax = 418 \pm 4\,\muHz$ and $\Dnu = 26.8 \pm 1.9\,\muHz$, which are consistent with our results (see Sec.~\ref{sec:power-spectra}).  

After the analysis for this paper had largely been completed, a second sector of TESS data for \baql\ became available (Sector 81). This sector does not overlap with the SONG observations presented in this paper so we have not included it, and we defer that analysis to a subsequent paper.

\subsection{SARG (2009)}
\label{sec:sarg}

\begin{figure}
\centering
\includegraphics[width=\hsize]{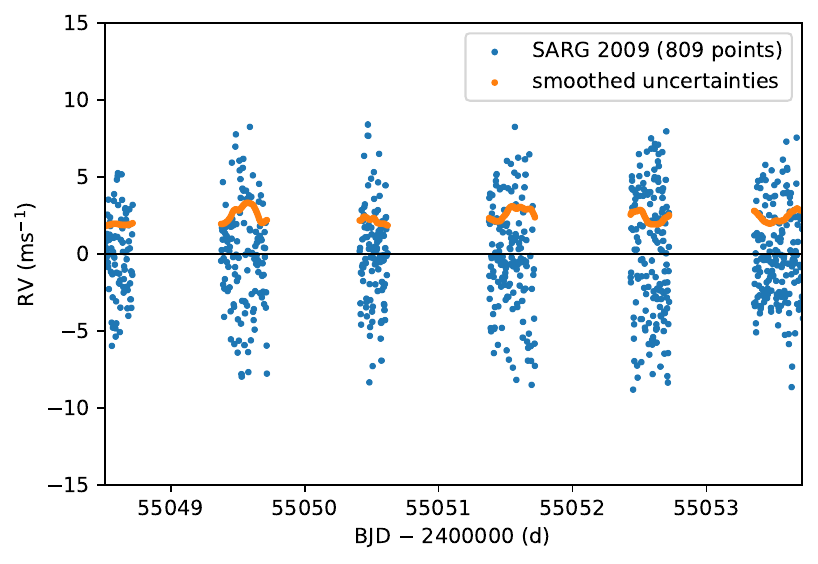}
\caption{Time series of RV measurements of \baql\ from SARG in 2009 (median uncertainty 2.3\,\ms; see Secs.~\ref{sec:sarg} and~\ref{sec:data-processing}).}
\label{fig:time-series-SARG}
\end{figure}

Solar-like oscillations were first detected in \baql\ by \citet{Corsaro++2012} using single-site observations with the SARG spectrograph at the Italian Telescopio Nazionale Galileo (TNG) 3.5~m telescope on La Palma. They observed over 6 nights in August 2009, using an iodine cell as the RV reference, and measured oscillations centred at $\numax=416\,\muHz$ (which corresponds to a period of 40\,min).  We have analysed those 809 RV measurements as part of this work (see Fig.~\ref{fig:time-series-SARG}).  They have a median sampling of 192\,s.

\subsection{HARPS (2008)}
\label{sec:harps}

\begin{figure}
\centering
\includegraphics[width=\hsize]{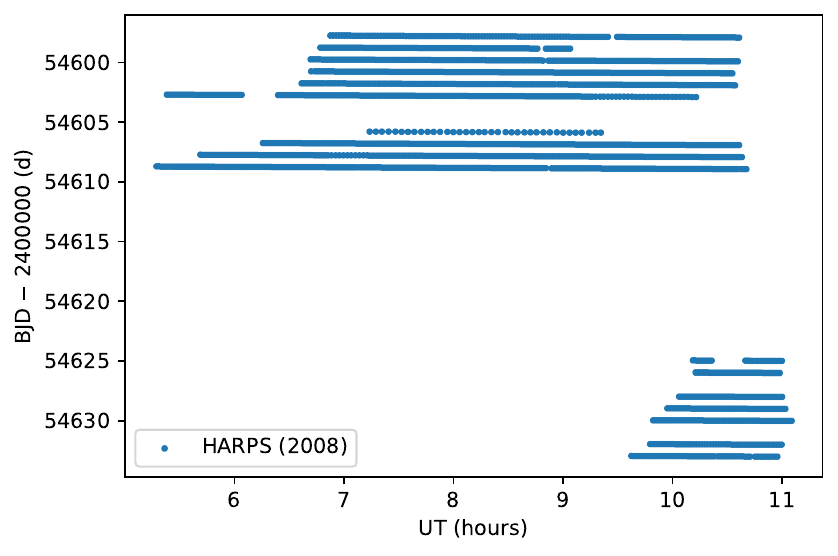}
\includegraphics[width=\hsize]{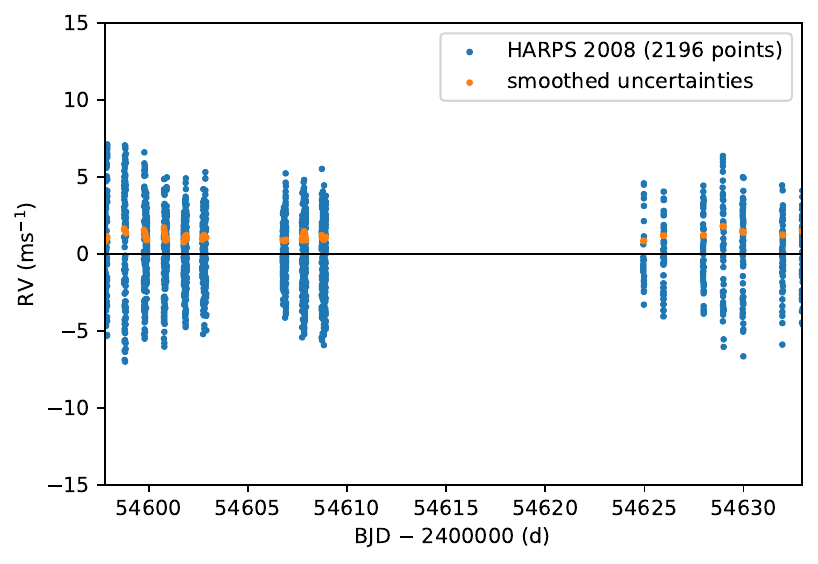}
\caption{Time series of \baql\ from HARPS in 2008.  Top: nightly coverage of the measurements.  Bottom: radial velocities (median uncertainty 1.1\,\ms; see Secs.~\ref{sec:harps} and~\ref{sec:data-processing}).}
\label{fig:time-series-HARPS}
\end{figure}

We retrieved 2200 unpublished RV measurements from the ESO archive that were obtained using the HARPS spectrograph (High Accuracy Radial velocity Planet Searcher) on the ESO 3.6m at La Silla in Chile. These measurements were taken on 17 nights spread over 35 nights in May and June 2008, with a median sampling cadence of 71\,s. As shown in the top panel of Fig.~\ref{fig:time-series-HARPS}, the observations on 8 nights spanned 4--5.5\,hr but the rest of the nights only contained 1--2\,hr of data.  The processed time series (see Sec.~\ref{sec:data-processing} for details) is shown in the bottom panel of Fig.~\ref{fig:time-series-HARPS}.

\section{Time-series processing}
\label{sec:data-processing}

\subsection{Introduction}
\label{sec:hpf}

Each of the time series described in Sec.~\ref{sec:observations} (SONG-Tenerife, SONG-Australia, TESS, SARG and HARPS) were processed separately using the steps described in the following sections (note that the SONG-Australia time series from the two 0.7-m telescopes were processed separately and then combined).  Some of these steps involved applying a high-pass filter to the time series to remove slow trends by subtracting a smoothed version of the time series from the original:
\begin{equation}
    y_{\rm hpf}(t_i) = y(t_i) - y_{\rm smoothed}(t_i). \label{eq:hpf}
\end{equation}
\new{Note that the subscripts $i$ and $j$ in Equations~\ref{eq:hpf}--\ref{eq:weights-normalization} are indices of individual points in the time series.}
To create $y_{\rm smoothed}$, we used a convolution in which each point was replaced by a weighted average of the points in its vicinity:
\begin{equation}
    y_{\rm smoothed}(t_i) = \frac{\sum_{j=1}^N y(t_i) w(t_i,t_j)}{\sum_{j=1}^N w(t_i,t_j)}, 
    \label{eq:ysmoothed}
\end{equation}
where $N$ is the total number of data points, and we chose the weighting function $w$ to be a super-Lorentzian:
\begin{equation}
    w(t_i,t_j) = \frac{1}{1 + \left(\frac{t_i-t_j}{\Delta t}\right)^4}.
    \label{eq:weights}
\end{equation}
The quantity $\Delta t$, which is the half-width at half-maximum of the weighting function, specifies the timescale of the high-pass filter. \new{The values of $\Delta t$ that we adopted are given in the next sections.}

\subsection{Removing bad data points}
\label{sec:bad-data}

Each time series contains some bad data points that needed to be identified and removed.  A small number of extreme outliers in the RV time series were removed by a simple clipping process (we used a threshold of $\pm20$\,\ms, which removed 1.8\% of the data points). 
We then identified less extreme outliers using the following procedure.

We first high-pass filtered the time series (see Sec.~\ref{sec:hpf}) but, in order to make this useful for identifying bad data points, we modified the smoothing process slightly.  Instead of using Eq.~\ref{eq:ysmoothed}, we created the smoothed version of the time series by replacing each data point by a weighted average of its neighbours {\em excluding the point itself}:
\begin{equation}
    y'_{\rm smoothed}(t_i) = \frac{\sum_{\substack{j=1\\j\ne i}}^N y(t_i) w(t_i,t_j)}{\sum_{\substack{j=1\\j\ne i}}^N w(t_i,t_j)}.
    \label{eq:ysmoothed-exclude}
\end{equation}
In the weighting function (Eq.~\ref{eq:weights}) we set $\Delta t$ to be the typical sampling time, calculated as the median value of $(t_{i+1}-t_i)$.  This small value of $\Delta t$ means that the high-pass filtering only left the scatter on very short time scales. 

    \begin{figure}
    \centering
    \includegraphics[width=\hsize]{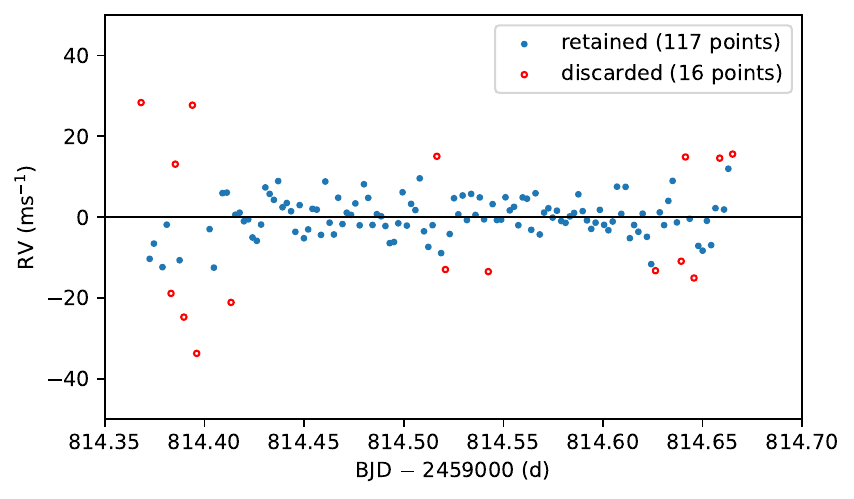}
    \caption{\new{One night of SONG-Tenerife data in 2022, illustrating bad data points that were discarded using the method outlined in Sec.~\ref{sec:bad-data}. Note that this night had an unusually large number of bad points.}}
    \label{fig:time-series-example}
    \end{figure}

The next step involved using this high-pass filtered time series, which we refer to as $y_1$, to identify bad data points.  
We defined these bad points as having $|y_1| > 5|y_{1, \rm smoothed}|$, where $|y_{1, \rm smoothed}|$ is the smoothed version of $|y_1|$ calculated based on Eq.~\ref{eq:ysmoothed} with a value of $\Delta t = 0.05$\,d (1.2\,h).
The result of this process was an observed time series in which bad data points have been identified and removed.  \new{To illustrate the process, an example is shown in Fig.~\ref{fig:time-series-example}, where we have chosen a night with an unusually large number of bad points.}
 
\subsection{Calculating uncertainties}
\label{sec:uncertainties}

The quality of the data varies during each time series and so it is very important to weight the data points when calculating the Fourier amplitude spectrum.  For this we need to allocate an uncertainty $\sigma(t_i)$ to each data point $y(t_i)$.  We did this using the time series $y_1$ described in Sec.~\ref{sec:bad-data}, recalculated with the bad data points removed.  The smoothed uncertainties were then calculated as follows:
\begin{equation}
    \sigma(t_i)^2 = \frac{\sum_{j=1}^N y_1(t_i)^2 w(t_i,t_j)}{\sum_{j=1}^N w(t_i,t_j)}, 
    \label{eq:sigma}
\end{equation}
\new{which measures the scatter on short timescales.}
The calculation uses the weighting function $w_\sigma$:
\begin{equation}
    w_\sigma(t_i,t_j) = \frac{1}{1 + \left(\frac{t_i-t_j}{\Delta t}\right)^8}
    \label{eq:weights-sigma}
\end{equation}
and the half-width was once again $\Delta t = 0.05$\,d (1.2\,h).  We used the exponent of 8 instead of 4 to make the function even more box-like, to give less sensitivity to points that were not close to the point under consideration.
In some cases, a small segment of the time series was given too much weight because it had very low scatter.  To avoid this, we set a lower limit on $\sigma(t_i)$ of 0.7 times the median value.

Before calculating the Fourier amplitude spectrum of the time series, we applied a high-pass filter (Eqs.~\ref{eq:hpf}--\ref{eq:weights}) with $\Delta t = 0.05$\,d (1.2\,h). 
These time series are shown as blue points in Figs.~\ref{fig:time-series-2022} to~\ref{fig:time-series-SARG}.  The amplitude spectrum was then calculated using $\sigma(t_i)^{-2}$ as weights \citep[see][]{Frandsen++1995} and the results are discussed in the next section.
Finally, we rescaled $\sigma(t_i)$ to reflect the mean noise in the weighted amplitude spectrum, $\sigma_{\rm amp}$, as measured at high frequencies.  That is, we multiplied the uncertainties by a constant so that they satisfied equation (3) of \citet{Butler++2004}:
\begin{equation}
    \sigma_{\rm amp}^2 \sum_{i=1}^N \sigma(t_i)^{-2} = \pi. 
    \label{eq:weights-normalization}
\end{equation}

We repeated the steps in Sections~\ref{sec:bad-data} and ~\ref{sec:uncertainties} until no additional data points were removed.  The total number of data points removed in each time series was 2--3\%.
The final uncertainties are shown as orange points in Figs.~\ref{fig:time-series-2022} to~\ref{fig:time-series-SARG} and the median values are given in the figure captions. We would be happy to provide these time series of measurements and uncertainties upon request.



\section{Analysis of oscillations}

\subsection{Weighted power spectra}
\label{sec:power-spectra}
\begin{figure*}
    \centering
    \includegraphics[width=0.8\hsize]{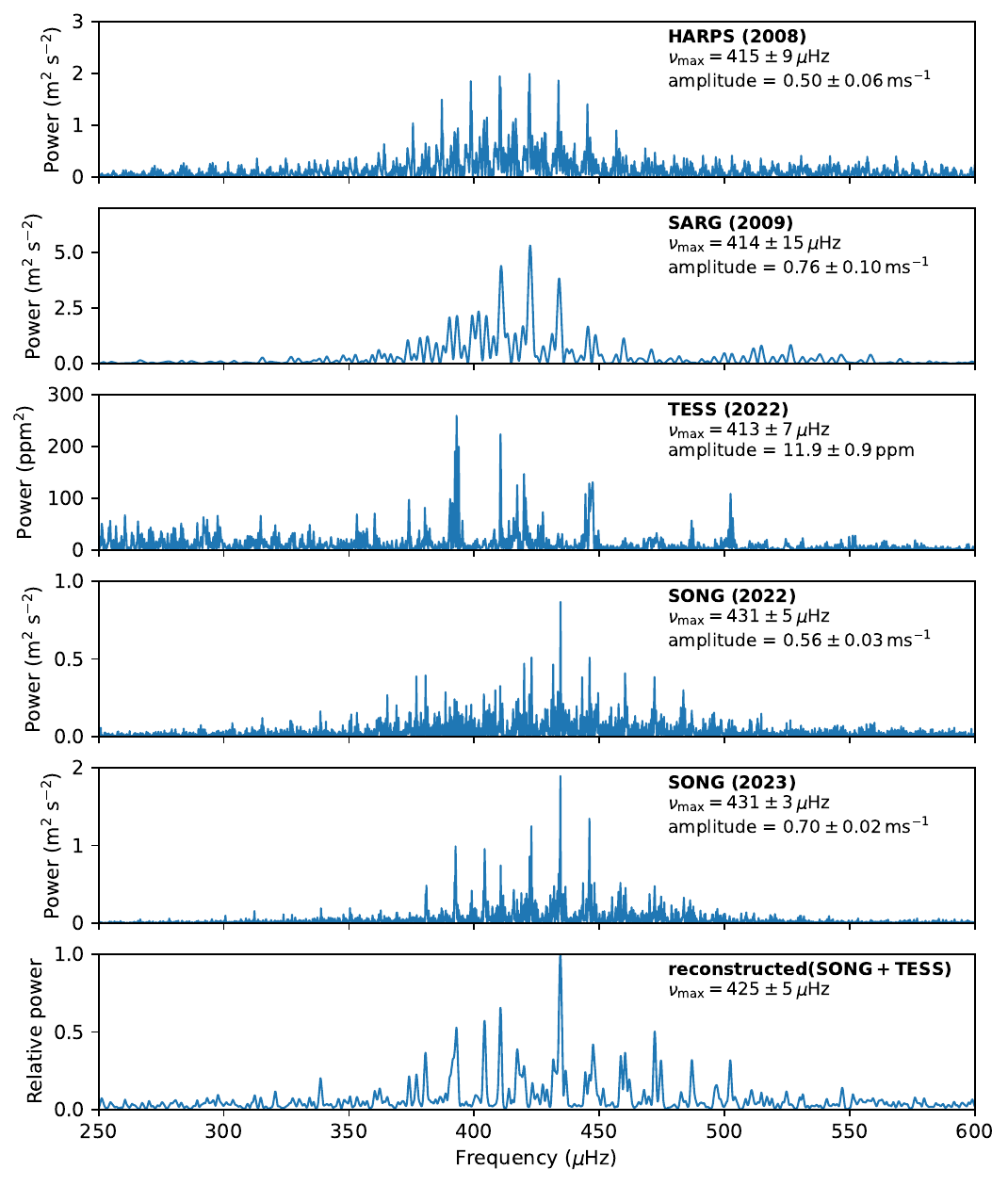}
    \caption{Power spectra of \baql\ showing solar-like oscillations from four time series in radial velocity and one in TESS photometry, with measurements of \numax\ and amplitude per radial mode (see Sec.~\ref{sec:power-spectra}). The bottom panel shows the reconstructed power spectrum from the combined SONG and TESS data, smoothed to a resolution of 1\,\muHz\ (see Sec.~\ref{sec:frequencies}).}
    \label{fig:power-spectra}
\end{figure*}

The weighted power spectra of the five time series discussed above are shown in Fig.~\ref{fig:power-spectra}.  The oscillations are clearly detected in all data sets, and the TESS data also show rising power towards low frequencies arising from surface granulation.  As mentioned in the Introduction, the granulation background is much lower (relative to the oscillations) in RV measurements.
For each power spectrum we indicate \numax\ and the amplitude per radial mode, which we measured using the method described by \citet{kjeldsen++2008-solar-amp}.  In summary, this involves heavily smoothing the power spectrum, converting to power density, fitting and subtracting the background from white noise and granulation, and multiplying by $c/\Dnu$, where $c$ is the effective number of modes per order (normalized to $\ell=0$).  We used $\Dnu=27.3\,\muHz$ (see \new{Sec~\ref{sec:frequencies}}), and chose $c=4.09$ for the RV measurements and $c=2.95$ for the TESS photometry (see \citealt{kjeldsen++2008-solar-amp} for more details).

The values of \numax\ show some scatter between the different data sets, which presumably reflects fluctuations arising from the stochastic nature of the excitation and damping.  \new{On the other hand, some of the scatter could reflect variations between the different observing techniques.}
There is perhaps some indication that the RV amplitude has changed during the 14 years spanned by the observations, which might be expected if the star has a magnetic activity cycle (see Sec.~\ref{sec:variations} for further discussion). 

These data sets give an excellent opportunity to compare measurements of oscillations in velocity and photometry, especially since they were obtained simultaneously. The best previous example of such a comparison comes from the Sun and we discuss solar measurements in the next section, before returning to \baql\ in Sec.~\ref{sec:phot-vel-ratio}.


\subsection{The solar oscillations in photometry and velocity}
\label{sec:solar}

\begin{figure}
    \centering
    \includegraphics[width=\hsize]{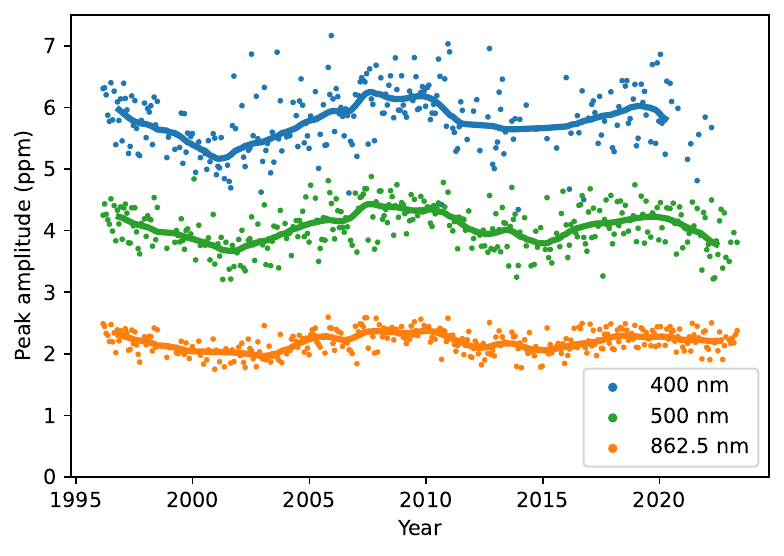}
    \caption{Peak amplitude of the solar oscillations measured over 27 years using data from the three-channel solar photometers (SPM) on the VIRGO Experiment aboard the SOHO spacecraft.  Points are measurements from 25-day non-overlapping segments of the time series and the solid curves result from a median filter followed by smoothing with a Gaussian window (FWHM = 1\,yr). \new{The lack of points in the 400-nm data towards the end of the series is a result of reduced data quality, with too few points for reliable amplitude measurements.}}
    \label{fig:solar-virgo-amp}
\end{figure}

We briefly digress to discuss the amplitude of oscillations in the Sun.
In photometry, the best available intensity measurements come from the VIRGO instrument (Variability of solar IRradiance and Gravity Oscillations) on the SOHO spacecraft (Solar \& Heliospheric Observatory), which has been operating since 1996 \citep{Frohlich++1995, Frohlich++1997}.  VIRGO provides photometric time series in three channels: blue (402\,nm), green (500\,nm) and red (862.5\,nm). We have measured the solar amplitude over the full data set by applying the method described in Sec.~\ref{sec:power-spectra} to 25-day non-overlapping segments of the time series. The results are shown in Fig.~\ref{fig:solar-virgo-amp}. 

It is clear from Fig.~\ref{fig:solar-virgo-amp} that the photometric amplitude of the solar oscillations varies over the Sun's 11-year activity \citep[see also][]{Jimenez-Reyes++2003, Kim+Chang2022} and the same behaviour is seen in velocity amplitudes \citep{Chaplin++2000, Chaplin++2003, kjeldsen++2008-solar-amp, Kiefer++2018, Howe++2022}. We also see that the photometric amplitude depends on wavelength, which is a well-known phenomenon for solar-like oscillations \citep[e.g.,][]{Kjeldsen+Bedding1995, Ballot++2011, Lund2019, Sreenivas++2025}. In order to compare photometric amplitudes from different instruments (and also to compare photometry with velocity), it is convenient to consider the {\em bolometric amplitude} of an oscillating star. This is the fractional variation in amplitude that would be measured if the observations covered all wavelengths. The notation for this quantity used by \citet{Kjeldsen+Bedding1995} was $\left(\delta L/L\right)_{\rm bol}$ but here we adopt the more compact notation of $A_{\rm bol}$ \citep[e.g.,][]{Lund2019}.  As noted by \citet{Kjeldsen+Bedding1995}, this is related to the amplitude at a given wavelength by the approximate relation
\begin{equation}
    A_{\rm bol} = A_\lambda \frac{\lambda}{\lambda_{\rm bol}}.
    \label{eq:Abol_approx}
\end{equation}
Here,
\begin{equation}
    \lambda_{\rm bol} = \frac{hc}{4k\Teff}
    \label{eq:lambda_bol}
\end{equation}
is the wavelength at which the observed luminosity amplitude ($A_\lambda$) is equal to~$A_{\rm bol}$. A more accurate version of Eq.~\ref{eq:Abol_approx} was derived by \citet{OToole++2003}.\footnote{Note Eq.~5 of \citet{OToole++2003} is missing a factor of 4 in the denominator. The correct version is given in Eq.~\ref{eq:lambda_bol} above.} Assuming the luminosity changes come entirely from changes in temperature (so that changes in stellar radius can be neglected), the relation is
\begin{equation}
    A_{\rm bol} = A_\lambda \frac{\lambda}{\lambda_{\rm bol}} \left(1 - e^{-4\lambda_{\rm bol}/\lambda}\right).
    \label{eq:Abol_exact}
\end{equation}

We used Eq.~\ref{eq:Abol_exact} to estimate the bolometric amplitude of the Sun from our time series of VIRGO measurements. We took the weighted mean of the green and red channels, since these fall on either side of the bolometric wavelength of the Sun, which is $\lambda_{\rm bol, \odot} = 623$\,nm.  We averaged the measurements over two solar cycles (22 yrs) to get
\begin{equation}
    A_{\rm bol, \odot} = 3.12 \pm 0.08\,{\rm ppm}.
    \label{eq:Abol_sun}
\end{equation}
This is slightly lower than the values previously measured from VIRGO data, namely
$3.58 \pm 0.16$\,ppm obtained by \citet{Michel++2009}%
\footnote{\citet{Michel++2009} measured $2.53 \pm 0.11$\,ppm, but this was the rms amplitude and needs to be multiplied by~$\sqrt{2}$.} 
and $3.5 \pm 0.2$\,ppm obtained by \citet{Huber++2011}.  
Given that the value in Eq.~\ref{eq:Abol_sun} has been averaged over two full solar cycles, we consider it to be the most reliable.

We now turn to the velocity amplitude of the solar oscillations. Here, we are interested in observations made using stellar techniques\new{, which involve measuring the radial velocity from a wide wavelength range that includes a large number of spectral lines}. As discussed in detail by \citet{kjeldsen++2008-solar-amp}, these give different amplitudes to the single-line RVs measured by dedicated helioseismology instruments such as BiSON and GOLF.
The longest solar data using stellar techniques was measured with SONG-Tenerife over 57 consecutive days in mid-2018 by \citet{Andersen++2019}. They reported the average amplitude of the strongest radial modes as $v_{\rm osc, \odot} = 0.166 \pm 0.004$\,\ms\ (measured using the same method as in this paper---see Sec.~\ref{sec:power-spectra}). We can adjust for the fact that the VIRGO amplitude during those observations was 3.3 per cent higher than the average over two activity cycles, which then gives the mean solar amplitude in velocity to be
\begin{equation}
    v_{\rm osc, \odot} = 0.160 \pm 0.004\,\ms.
    \label{eq:vosc_sun}
\end{equation}
To conclude, Eqs.~\ref{eq:Abol_sun} and~\ref{eq:vosc_sun} give our best estimates for the mean oscillation amplitude in the Sun in photometry and velocity. 

\subsection{Photometry-to-velocity amplitude ratio of \baql}
\label{sec:phot-vel-ratio}

\baql\ now joins a small number of stars for which solar-like oscillations have been observed in both RV and photometry, and an even smaller number for which these observations were simultaneous (Procyon is the only published example of which we are aware; see \citealt{Huber++2011-procyon}).  This allows us to measure the ratio between the two types of measurements  \citep[see also][]{Houdek2010, Arentoft++2019, Zhou-Yixiao++2021}.  

\new{Note that simultaneous measurements are highly preferable for this purpose because of the large variations in amplitude that are intrinsic to solar-like oscillations.  }
We measured the RV amplitude in the part of the SONG time series that overlapped with the TESS data (see Fig.~\ref{fig:time-series-2022-nightly}) to be 
$v_{\rm osc} = 0.497 \pm 0.044$\,\ms. 
Meanwhile, the TESS photometric amplitude was $11.9 \pm 0.9$\,ppm (Fig.~\ref{fig:power-spectra}).  Because TESS has a very broad passband, we did not use Eq.~\ref{eq:Abol_exact}. Instead, we converted the TESS amplitude to the bolometric amplitude by multiplying by $c_{T-\text{bol}} = 1.066\pm 0.011$, which we calculated from Eq.~7 of \citet{Lund2019} using $\Teff = 5092 \pm 50$\,K. This gave the photometric amplitude of \baql\ as
$A_{\rm bol} = 12.7 \pm 1.0$\,ppm.  

The photometry-to-velocity ratio for \baql\ is therefore
\begin{equation}
    A_{\rm bol}/v_{\rm osc} = 26.6 \pm 3.1\,\rm ppm/\ms.  
    \label{eq:amp-ratio-baql}
\end{equation}
The same ratio for the Sun, using Eqs.~\ref{eq:Abol_sun} and~\ref{eq:vosc_sun}, is
\begin{equation}
    A_{\rm bol,\odot}/v_{\rm osc,\odot} = 19.5 \pm 0.7\,\rm ppm/\ms.   
    \label{eq:amp-ratio-sun}
\end{equation}
According to \citet{Kjeldsen+Bedding1995}, this ratio is expected to scale inversely with \Teff.  The effective temperatures of the Sun and \baql\ are in the ratio $1.134\pm0.011$, whereas the values in Eqs.~\ref{eq:amp-ratio-baql} and~\ref{eq:amp-ratio-sun} are in the ratio $1.36\pm0.17$. These values agree within the uncertainties ($1.3\sigma$), although it would clearly be desirable to improve the accuracy for \baql. 

\begin{figure}
\centering
\includegraphics[width=\hsize]{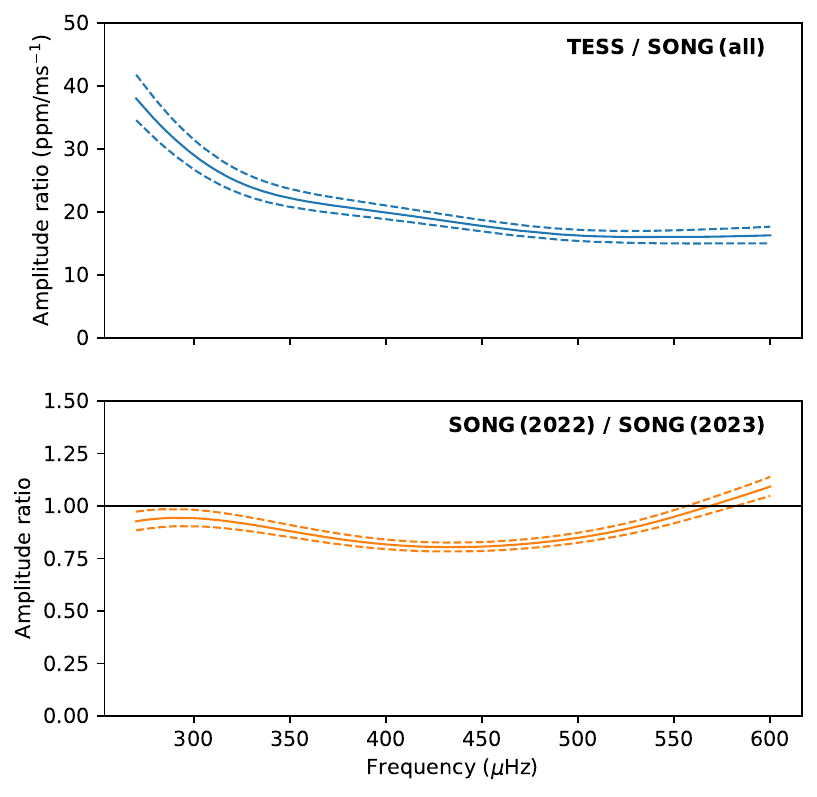}
\caption{Ratio between amplitude spectra obtained with TESS and SONG (upper panel) and between the two seasons of SONG data (lower panel), smoothed to $4\Dnu$ (see Sec.~\ref{sec:phot-vel-ratio}).
}
\label{fig:amp-ratio}
\end{figure}

We have also examined the photometry-to-velocity ratio as a function of frequency, as shown in the upper panel of Fig.~\ref{fig:amp-ratio}.  This plot was created by dividing a heavily smoothed version of the TESS power spectrum by a similarly smoothed version of the SONG power spectrum (using all data) and then taking the square root.  For both TESS and SONG, we first removed the background by fitting components for white noise and granulation (where the latter used a simple Harvey profile; see \citealt{Harvey1988}). The smoothing was achieved by convolving with a Gaussian having a FWHM of $4\Dnu$.
We see a strong trend in the ratio as a function of frequency which does not appear to agree with theoretical expectations \citep{Houdek2010} and clearly deserves further study.  
\new{The granulation background, which is stronger in TESS data, has been subtracted before calculating the amplitude ratio but we note that incorrect subtraction would lead to trends at in amplitude at low frequencies.  For this reason, we restricted the plot in Fig.~\ref{fig:amp-ratio} to frequencies above 270\,\muHz, where the granulation background is very low.}
As a further check on this result, we also calculated the ratio between the two sets of SONG RV data.  As shown in the lower panel of Fig.~\ref{fig:amp-ratio}, this ratio is independent of frequency but shows a possible increase in the oscillation amplitude from 2022 to 2023 (see Sec.~\ref{sec:variations}).

Regarding the increase in the photometry-to-velocity ratio towards lower frequencies, we note that a similar trend was reported in the Sun by \citet{Jimenez2002}. By analyzing the heights of individual modes as measured by VIRGO and GOLF, they found the amplitude ratio at 2\,mHz to be higher by a factor of two higher than at 3--4\,mHz (see Figure 6 in that paper). Magnetic activity probably affects amplitudes and phases, which theoretical models do not include. We note that the value of \numax\ would also be affected, causing it to be lower when measured in photometry than in velocity. Indeed, this effect may be be partially responsible for such a finding in the K0 dwarf $\sigma$~Dra by \citet{Hon++2024}, using photometry from TESS and radial velocities from the Keck Planet Finder (KPF). 

\subsection{Comparing the simultaneous intensity and velocity signals}
\label{sec:overlap}

Since a small part of the SONG data overlaps in time with TESS, we can directly compare the signals in intensity and velocity. To do this, we isolated the oscillation signal in each time series using a band-pass filter. This was implemented by using the standard method of iterative sine-wave fitting (see also Sec.~\ref{sec:frequencies}) to extract 10,000 sinusoidal components in the frequency range 330--550\,\muHz, \new{which covers the region in which we see excess power from oscillations}.  We then reconstructed the time series as the sum sinusoids using these extracted frequencies, amplitudes and phases.

\begin{figure*}
    \centering
    \includegraphics[width=0.8\hsize]{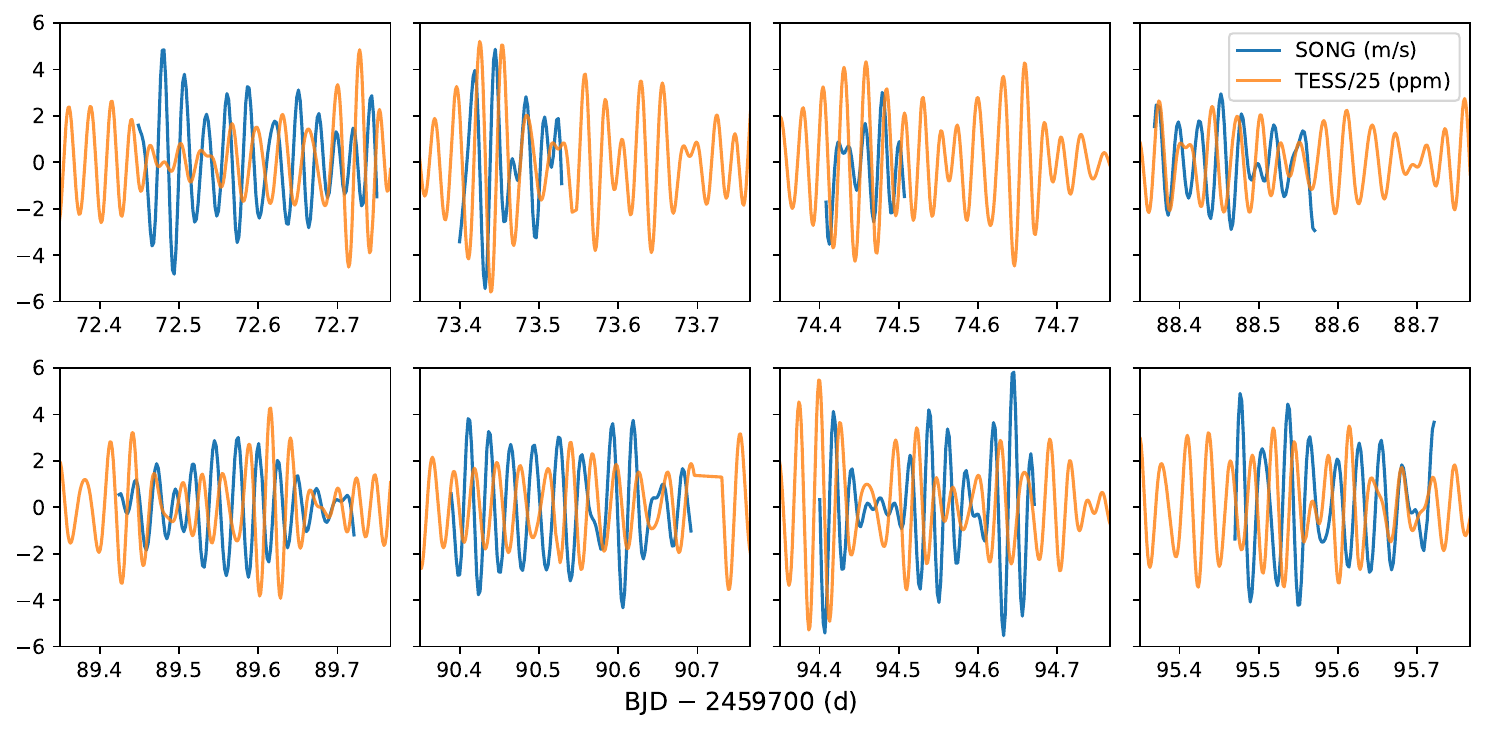}
    \caption{Observations of \baql\ on eight nights that had overlap between SONG and TESS. Each panel shows a 10-hr segment, with radial velocities from SONG-Tenerife in \ms\ (blue curves) and TESS photometry divided by 25\,ppm (orange curves).  Both series have been bandpass-filtered to show variations in the range 330--550\,\muHz. }
    \label{fig:overlap}
    \end{figure*}

Figure~\ref{fig:overlap} shows the results for the 8 nights on which SONG was observing simultaneously with TESS. 
We see a clear indication of a phase shift between the SONG and TESS data. The signal-to-noise and the length of the time series are not high enough to measure the phase shifts for
individual oscillation modes, so we estimated the phase shift for the combined oscillation signal. 
We initially did this by calculating the cross correlation between the TESS and SONG data for each of the segments separately. This gave a shift in time between the two series for each segment, which we converted to a phase shift using the typical timescale of the oscillations \new{(measured from the peak of the cross-correlation function for that segment)}.
Only five of the segments were long enough to provide a robust measurement, and we calculated the average phase shift to be $-151^{\circ} \pm 25^{\circ} = -2.64 \pm 0.44$\,rad. 

As well as treating the overlapping segments separately, we also calculated the cross correlation using the entire section for which there was overlap. This gave better accuracy and we calculated the phase shift to be $-113^{\circ} \pm 7^{\circ}$. In other words, the peak of the SONG oscillations (highest positive RV) occurs 31 per cent of an oscillation period before the peak of the TESS signal.

In the Sun, the corresponding phase shift can be measured using SOHO observations \citep{Jimenez2002}. We used one month of data from GOLF and VIRGO to measure the phase shift in the Sun to be $-125.6^{\circ} \pm 0.5^{\circ}$. This is close to the \baql\ value but differs by $1.8\sigma$ \new{(although we do not necessarily expect exact agreement)} and a longer overlapping time series for \baql{} is needed to give a more precise measurement. We also note that in the Sun, \citet{Jimenez2002} found the phase difference between velocity and intensity to vary by as much as 10 degrees over the solar activity cycle (their Figure 4).  They also noted that the effects of changing magnetic activity are not included in the models by \citet{Houdek++1995}.

Importantly, these results indicate that high-cadence photometric observations with TESS could be used to infer the corresponding velocity oscillation signal, and hence to improve the precision of simultaneous RV measurements.  This is highly relevant to high-precision RV searches for exoplanets \citep[e.g.,][]{Dumusque++2011, Yu-Jie++2018, Chaplin++2019, Nieto_Diaz2023, Zhao-Lily2023, Luhn++2023, Gupta+Bedell2024, OSullivan+Aigrain2024, Zhao-Yinan++2024, Beard++2025, Tang++2025}.

It is also interesting to compare our measured phase difference with 3-D hydrodynamical simulations of stellar atmospheres. As a preliminary comparison, we have used a simulation that was previously computed using the {\sc stagger} code for a star with $\Teff = 5000$\,K and $\log g = 3.5$ \citep{Magic++2013}.  These values are similar to those of \baql\ ($\Teff = 5092$\,K and $\log g = 3.55$). The simulated time series consists of 3000 snapshots with a sampling cadence of 166\,s and a total duration of 138.33 hours of stellar time. We measured the phase difference between the stellar flux variations and the vertical fluid velocity to be $-121^\circ \pm 6^\circ$, which is in excellent agreement with our measurements for \baql.

\subsection{Extraction of mode frequencies}
\label{sec:frequencies}

To extract frequencies of the oscillation modes, we first generated a reconstructed power spectrum based on the data from SONG (2022), SONG (2023) and TESS.  The SONG data, in particular, are affected by gaps in the time series that introduce sidelobes in the power spectra.  The first step was to reconstruct power spectra for each data set using the standard method of iterative sine-wave fitting (also called prewhitening), using the uncertainties as weights.  For each of these three data sets, we extracted 10,000 sinusoidal components in the frequency range 200--700\,\muHz\ to ensure that all power---including the noise---was included.  We then created a reconstructed power spectrum as the sum of $\delta$ functions with those frequencies and heights.  

Having done this, we corrected the TESS spectrum for the granulation background, which rises towards lower frequencies.  We did this by fitting and dividing by a function of the form $(\nu/\numax)^{-2}$, which matches the slope of the granulation background \citep[e.g.,][]{Sreenivas++2024}.
Since SONG and TESS observe different quantities, we normalized each reconstructed power spectrum to have a mean of 1 before averaging them. 
This final reconstructed power spectrum was smoothed by convolving with a Gaussian with a FWHM of 1\,\muHz, chosen to match approximately the lifetime of the modes. The reconstructed spectrum is shown in the bottom panel of Fig.~\ref{fig:power-spectra}.  

\begin{figure}
    \centering
    \includegraphics[width=\hsize]{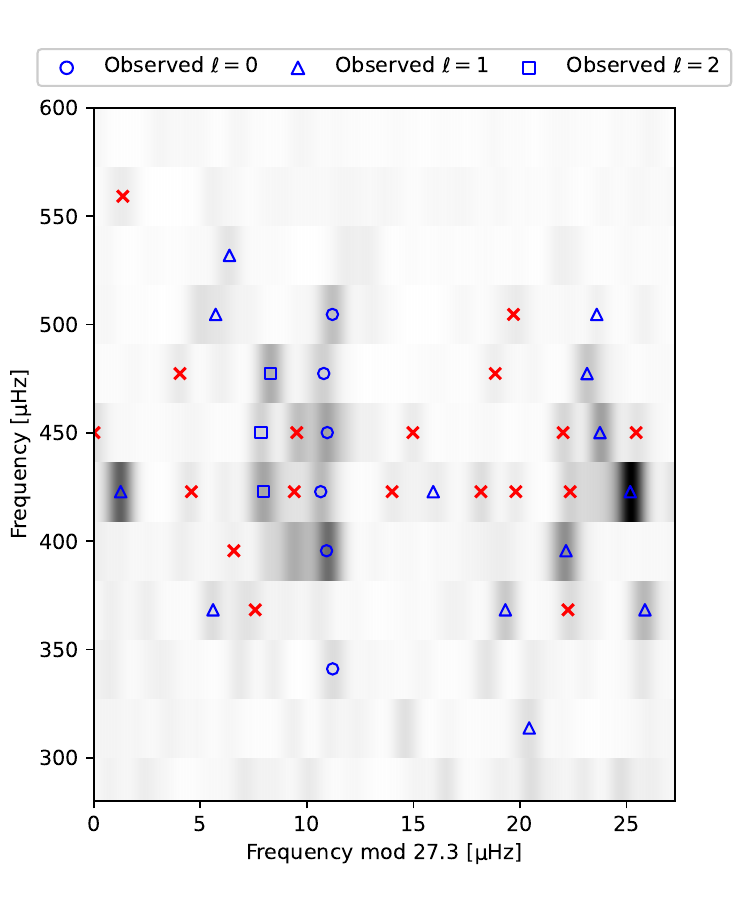}
    \caption{Power spectrum of \baql\ reconstructed from the SONG and TESS data, shown in \echelle\ format.  The greyscale shows the reconstructed power spectrum (Sec.~\ref{sec:frequencies}) and the open blue symbols show the identified modes, while red crosses are assumed to be noise peaks (see Table~\ref{tab:freqs}).}
    \label{fig:echelle-reconstructed}
    \end{figure}

The reconstructed spectrum is shown in \echelle\ format in Fig.~\ref{fig:echelle-reconstructed}, where the power spectrum has been divided into equal segments of width \Dnu\ that are stacked vertically. The value of $\Dnu = 27.3\,\muHz$ was chosen to make the ridges vertical. This \echelle\ diagram reveals a pattern typical of a star in this evolutionary state, with a clear vertical ridge corresponding to modes with $\ell=0$, accompanied by a weaker ridge to its left ($\ell=2$ modes) and several other peaks that are spread over the rest of the diagram (mixed $\ell=1$ modes).  Very similar patterns have been seen in late subgiants and early red giant branch (RGB) stars observed by \kepler, particularly in KIC~4351319 \citep[nicknamed `Pooh';][]{di-Mauro++2011} and KIC~7341231 \citep[`Otto';][]{Deheuvels++2012}.

We extracted oscillation mode frequencies from the reconstructed power spectrum by measuring the strongest peaks in the frequency range 300--550\,\muHz, which is the region that contains the oscillations.  The 40 strongest peaks are listed in Table~\ref{tab:freqs}.
There is always a risk that the pre-whitening could latch on to a sidelobe, and so we must keep in mind that some modes maybe be off by one cycle per day ($\pm 11.57\,\muHz$). As a check, we applied the Gold deconvolution algorithm \citep{Morhac++2003, Li-Yaguang++2025} and found very similar results. 

\begin{table}
\caption{Extracted frequencies for \baql, ordered by increasing uncertainty (see Sec.~\ref{sec:frequencies}). Note that frequencies have not been corrected for the radial velocity of the star. The identification of noise peaks is discussed in Sec.~\ref{sec:identification}. }
\label{tab:freqs}
\centering
\begin{tabular}{ccc}
\hline\hline
Rank & Frequency (\muHz) & Identification\\
\hline
 1  & $434.44 \pm 0.22$ &  $\ell=1$ \\
 2  & $410.52 \pm 0.26$ &  $\ell=1$ \\
 3  & $404.14 \pm 0.28$ &  $\ell=1$ \\
 4  & $392.90 \pm 0.29$ &  $\ell=0$ \\
 5  & $472.13 \pm 0.29$ &  $\ell=2$ \\
 6  & $447.50 \pm 0.31$ &  $\ell=0$ \\
 7  & $417.23 \pm 0.32$ &  $\ell=2$ \\
 8  & $380.56 \pm 0.33$ &  $\ell=1$ \\
 9  & $460.30 \pm 0.33$ &  $\ell=1$ \\
10  & $458.57 \pm 0.34$ &  noise \\
11  & $431.62 \pm 0.34$ &  noise \\
12  & $474.62 \pm 0.35$ &  $\ell=0$ \\
13  & $486.98 \pm 0.35$ &  $\ell=1$ \\
14  & $502.31 \pm 0.35$ &  $\ell=0$ \\
15  & $419.91 \pm 0.36$ &  $\ell=0$ \\
16  & $436.55 \pm 0.38$ &  noise \\
17  & $444.39 \pm 0.38$ &  $\ell=2$ \\
18  & $376.95 \pm 0.39$ &  noise \\
19  & $418.67 \pm 0.39$ &  noise \\
20  & $374.01 \pm 0.40$ &  $\ell=1$ \\
21  & $446.07 \pm 0.40$ &  noise \\
22  & $338.62 \pm 0.41$ &  $\ell=0$ \\
23  & $462.00 \pm 0.42$ &  noise \\
24  & $423.26 \pm 0.44$ &  noise \\
25  & $496.83 \pm 0.45$ &  $\ell=1$ \\
26  & $427.43 \pm 0.45$ &  noise \\
27  & $547.04 \pm 0.47$ &  noise \\
28  & $362.27 \pm 0.48$ &  noise \\
29  & $413.84 \pm 0.48$ &  noise \\
30  & $451.52 \pm 0.48$ &  noise \\
31  & $429.07 \pm 0.48$ &  noise \\
32  & $510.81 \pm 0.49$ &  noise \\
33  & $467.87 \pm 0.49$ &  noise \\
34  & $360.29 \pm 0.50$ &  $\ell=1$ \\
35  & $320.56 \pm 0.50$ &  $\ell=1$ \\
36  & $524.76 \pm 0.50$ &  $\ell=1$ \\
37  & $482.67 \pm 0.51$ &  noise \\
38  & $514.72 \pm 0.52$ &  $\ell=1$ \\
39  & $388.55 \pm 0.53$ &  noise \\
40  & $425.19 \pm 0.53$ &  $\ell=1$ \\
\hline
\end{tabular}
\end{table}

In order to estimate uncertainties on the frequencies, we ran a number of simulations for modes with a lifetime similar to the Sun. We generated time series with the same length as the data and smoothed the power spectra to a resolution of 1\,\muHz\ before measuring the peak positions. We ran simulations at different SNR to estimate the error bars for the individual extracted modes \citep[see][]{de-ridder++2006}. 

The mode identifications are listed in Table~\ref{tab:freqs}. We identified 6 radial modes ($\ell=0$) and 3 quadrupolar modes ($\ell=2$) based on their positions in the \echelle\ diagram (Fig.~\ref{fig:echelle-reconstructed}). We also identified 13 mixed $\ell=1$ modes, using the method described in the next section. Finally, 18 peaks in Table~\ref{tab:freqs} that were not identified as modes are presumably noise peaks (red crosses Fig.~\ref{fig:echelle-reconstructed}).

We note that the radial velocity of \baql\ is $v_r = -41.50 \pm 0.68$\,k\ms\ \citep{Steinmetz++2020}, which means that the oscillation frequencies have been Doppler shifted by a factor of $(1+v_r/c)$, where $c$ is the speed of light. The observed frequencies should be multiplied by this factor before carrying out detailed modelling \citep{Davies++2014}, which amounts to a shift of 0.047--0.065\,\muHz\ over the range of detected modes.  We show the uncorrected frequencies in Table~\ref{tab:freqs} but made the Doppler correction before carrying out the modelling described in Sec.~\ref{sec:models}.

\subsection{Identifying mixed modes}
\label{sec:identification}

While it is straightforward to identify $\ell=0$ and~2 modes using the \echelle\ diagram (Fig.~\ref{fig:echelle-reconstructed}), it is more difficult to decide which of the remaining peaks represent $\ell=1$ modes and which are noise peaks. To do this, we took advantage of the fact that $\ell=1$ mixed modes follow a definite pattern. They result from coupling of pure p~modes and pure g~modes, both of which are in the asymptotic regime. This means pure p~modes with radial order $n_p$ are approximately equally spaced \new{in frequency} by \Dnu, with $\nu_p\approx\Dnu(n_p+\epsilon_{p,\ \ell=1})$. Here, $\epsilon_{p,\ \ell=1}$ is the p-mode phase term. Similarly, pure g~modes are approximately equally spaced \new{in period} by $\Delta\Pi_1$, with $\new{1/}\nu_g\approx\Delta\Pi_1(n_g+\epsilon_g)$, where $n_g$ is the radial order of each g~mode and $\epsilon_g$ is the g-mode phase term.

\begin{figure}
\centering
\includegraphics[width=\hsize]{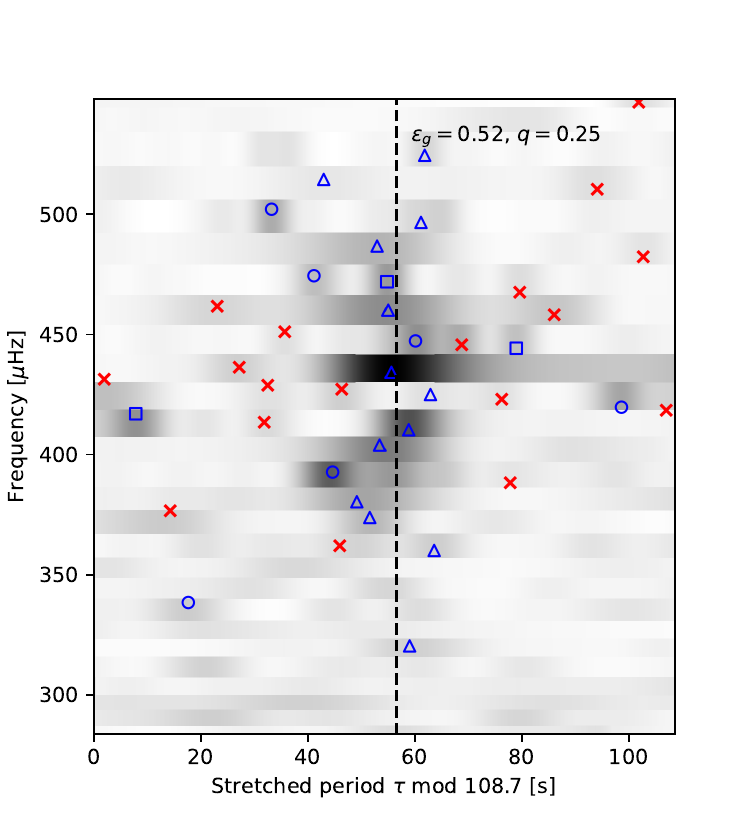}
\caption{Stretched period \echelle\ diagram, as described in Sec.~\ref{sec:identification}. The greyscale shows the reconstructed power spectrum, and the symbols showing identified modes (blue) and noise peaks (red crosses) have the same meanings as in Fig.~\ref{fig:echelle-reconstructed}. }
\label{fig:period-echelle}
\end{figure}

Initial modelling based on the $\ell=0$ and $\ell=2$ modes (see Sec.~\ref{sec:models} for a full description of the models) found good agreement with a model with a period spacing of $\Delta \Pi_1$ = 108.7~s. We used this value to construct a so-called stretched period \echelle\ diagram for the power spectrum, by converting frequency $\nu$ to stretched period $\tau$, \new{which is defined} as follows \citep{Mosser2015,Ong2023}:
\begin{equation}
    \tau = \frac{1}{\nu} + \frac{\Delta\Pi_1}{\pi} \arctan \left[ \frac{q}{\tan(\nu/\Delta\nu - \epsilon_{p,\ \ell=1})} \right].
\end{equation}
The dipole ($\ell=1$) mixed modes will align vertically in the stretched period \'echelle diagram, provided the asymptotic parameters appearing in the equation (the phase term $\epsilon_p$ and the coupling strength $q$) are appropriately chosen.
By iteratively optimising these parameters, we found that values of $\epsilon_{p,\ \ell=1}=1.95$ and $q=\new{0.25}$ produced \new{the best alignment, with the $l=1$ modes (blue triangles) reasonably close to} the vertical dashed line in Fig.~\ref{fig:period-echelle}. The position of that vertical alignment indicates that the pure-g modes are asymptotically spaced, with a phase of $\epsilon_g=\new{0.52}$. Based on their proximity to this line we identified 13 $\ell=1$ mixed modes, as listed in Table~\ref{tab:freqs}.

\subsection{Amplitude and frequency variations from 2022 to 2023}
\label{sec:variations}

Given that \baql\ appears to undergo a magnetic activity cycle with a period of $4.7 \pm 0.4$\,yr (Fig.~\ref{fig:S-index}), we might expect to see changes in the amplitudes and frequencies of its oscillation modes. Such changes are well-documented in the Sun \citep[see review by][]{Chaplin2014} and have also been seen in other stars \citep[e.g.,][]{Garcia++2010, Kiefer++2017, Salabert++2018, Santos++2018}.  With SONG we see an increase in oscillation amplitude from 2022 to 2023 of $25\% \pm 8\%$, which is a 3-$\sigma$ change (see Figs.~\ref{fig:power-spectra} and~\ref{fig:amp-ratio}). By comparison, the intrinsic variation of the oscillation amplitudes in the Sun over the solar cycle is 10\% peak-to-peak \citep{kjeldsen++2008-solar-amp}.
Continued SONG observations of \baql\ over the coming years should help to clarify whether these changes are related to its activity cycle.

We also compared the frequencies of individual modes in the SONG 2022 and 2023 data. The mean frequency difference is $0.019 \pm 0.082\,\muHz$ (a fractional change of $(0.4 \pm 1.9)\times 10^{-4}$), which is a non-detection.
The peak-to-peak shift in the solar frequencies over the full activity cycle is about 0.4\,\muHz\ (a fractional change of $1.3\times 10^{-4}$ \citep{Jimenez-Reyes++1998, Chaplin++2007, Chaplin2014}.  Therefore, our current upper limit on frequency shifts in \baql---when expressed as a fractional change---is greater than that seen in the Sun.

\section{Asteroseismic Modelling}
\label{sec:models}

\subsection{Modelling methods}
To perform asteroseismic model fitting, we constructed theoretical stellar models using MESA \citep[version r23.01.1;][]{paxton++2011-mesa, paxton++2013-mesa, paxton++2015-mesa, paxton++2018-mesa, paxton++2019-mesa, jermyn++2023-mesa} and GYRE \citep[version 7.0;][]{townsend+2013-gyre}. We used \texttt{pp\_and\_cno\_extras} for nuclear reaction rates \citep{jinareaclib2010,nacre1999}. The metallicity scale and metal mixture were chosen to be the AGSS09 solar abundance scale \citep{agss09}. Compatible opacity tables were used \citep[\texttt{`a09'}, \texttt{`lowT\_fa05\_a09p'}, and \texttt{`a09\_co'};][]{opal1996,Ferguson2005}.
The mixing-length theory was used for treating convection following the formalism of \citet{henyey++1964-mlt}. The grey Eddington $T$--$\tau$ relation was used as the atmospheric boundary condition \citep{eddington1926}. We did not include any mass loss, diffusion, overshoot, or other non-standard mixing processes. \new{In particular, we note that the extent of core overshooting is small at 1.2\Msolar\ \citep{Claret2018,Lindsay2024}.} All other settings adhered to MESA defaults. Our MESA inlists are available at this URL\footnote{\url{https://github.com/parallelpro/mesa_beta_aql}}.

The stellar model grid was sampled uniformly in a four-dimensional parameter space. The stellar mass ranged from $1.0$ to $1.4$ $M_\odot$ with a step size of $0.02$ $M_\odot$; the mixing length parameter $\alpha_{\rm MLT}$ ranged from 1.5 to 2.7 with a step size of 0.2 \new{(which comfortably includes the expected range; see Sec.~\ref{sec:model-parameters})}; the initial helium abundance $Y_{\rm init}$ ranged from 0.18 to 0.34 with a step size of 0.02; and the metallicity [M/H] ranged from -0.65 to 0.05 dex with a step size of 0.05 dex.
All tracks were evolved until $\Delta\nu$ dropped below 22~$\mu$Hz on the red giant branch. Stellar oscillation modes were calculated \new{using GYRE on-the-fly \citep{Bellinger2022, Joyce++2024}} for models that fell within a 5-$\sigma$ constraint box specified by $T_{\rm eff}$, $R$, and [M/H].

To evaluate each model in the grid, we constructed a $\chi^2$ function incorporating both the classical observables of \baql\ ($c=\{R, T_{\text{eff}}, {\text{[Fe/H]}}\}$; Table~\ref{tab:properties}) and the oscillation frequencies ($\nu=\{\nu_k; k=1,...N_l, l=0,1,2\}$; Table~\ref{tab:freqs}). The $\chi^2$ function was defined as
\begin{equation}\label{eq:chi2}
    \chi^2 = \sum_c\left(\frac{c_{\rm obs} - c_{\rm mod}}{\sigma_c} \right)^2 + \sum_\ell \frac{1}{N_\ell}\sum_k\left(\frac{\nu_{k, \rm obs} - \nu_{k, \rm mod}}{\sigma_{\nu_{k}}} \right)^2,
\end{equation}
\new{where the subscripts ``obs'' and ``mod'' indicate observational and modelling quantities, respectively, and the $\sigma$ values are the observational uncertainties.}
For each model $i$, the posterior probability was computed as $p_i \propto \Delta t_i\mathcal{L}_i = \Delta t_i\exp(-\chi_i^2/2)$, where $\Delta t_i$ is the time-step between adjacent models on the evolutionary track to account for numerical unevenness in the spacing.
Each stellar parameter $\theta_i$ was estimated from the cumulative distribution of $p_i(\theta_i)$, with the 50\% percentile representing the best-fitting value and the 16\% and 84\% percentiles providing the 1-$\sigma$ confidence intervals.

It is well established that model frequencies need to be corrected for inaccurate treatment of the near-surface layers \citep[e.g.,][]{Christensen-Dalsgaard++1996, Kjeldsen++2008-near-surface, Ball_2014, Jorgensen++2020, Belkacem++2021, Ong_2021, Li-Yaguang++2023}. In this work, we parametrized the surface correction using the cubic formula proposed by \citet{Ball_2014}. This formula can yield inaccurate results for mixed modes in red giants \citep{Ball++2018, Li-Tanda++2018}, but studies by \citet{Ong_2021} demonstrated its suitability for early subgiants such as \baql, which have high coupling strengths. 

\begin{figure}
    \centering
    \includegraphics[width=\hsize]{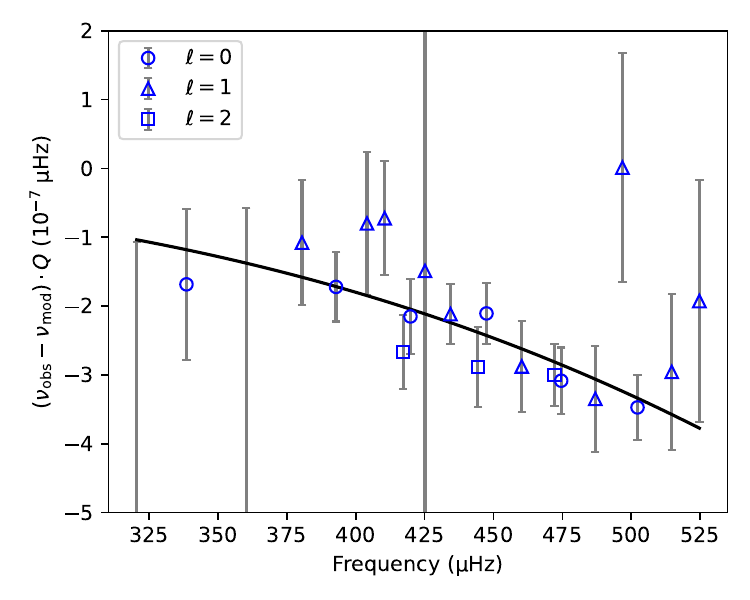}
    \caption{\new{Differences, scaled by mode inertia,  between the  observed frequencies and those of the best-fitting model (before making the surface correction). The differences approximately follow the cubic formula proposed by \citet{Ball_2014}, as shown by the solid line. }}
    \label{fig:freq-res}
\end{figure}

\new{Figure~\ref{fig:freq-res} shows the differences, scaled by mode inertia, between observed frequencies and those of the best-fitting model (before making the surface correction). We see that these scaled differences follow the cubic formulation within statistical uncertainties. Further, to avoid the surface correction from become unphysically large or small, we constrained it to vary smoothly as a function of stellar parameters \citep[e.g.,][]{Compton++2018, Li-Yaguang++2023, Reyes++2025}. This was done by setting} the amount of surface correction at \numax{} to be 4.5\% of \Dnu, which we determined based on the best-fitting model when we treated surface correction with free parameters, as originally proposed by \citet{Ball_2014}.
We note that this amount of surface correction is in agreement with that expected from the ensemble study by \citet[][their Equation 4 and Table 2]{Li-Yaguang++2023}.

\begin{figure}
\centering
\includegraphics[width=\hsize]{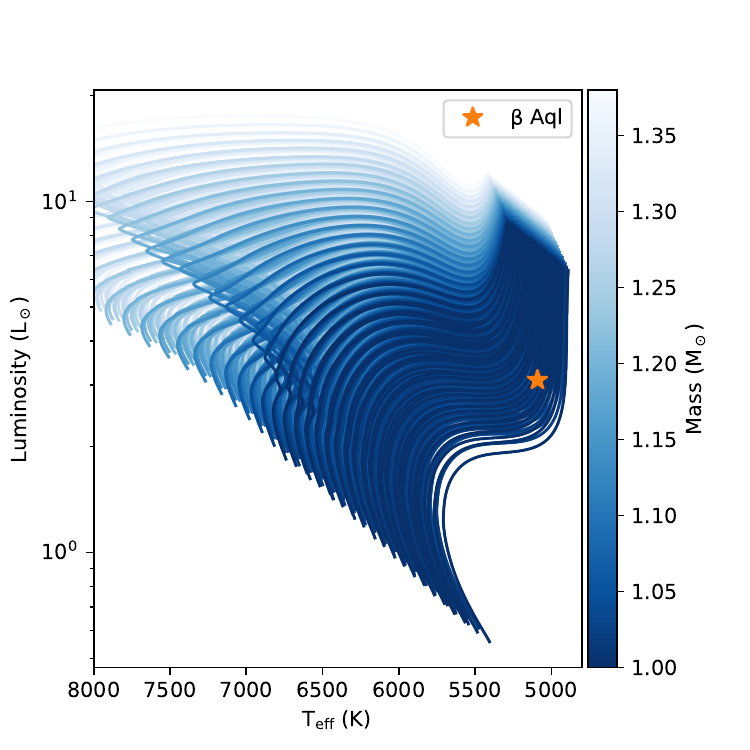}
\caption{\new{Hertzsprung--Russell diagram showing the observed location of \baql\ and the grid of theoretical models (see Sec.~\ref{sec:models}).} }
\label{fig:hrd}
\end{figure}

\begin{figure}
\centering
\includegraphics[width=\hsize]{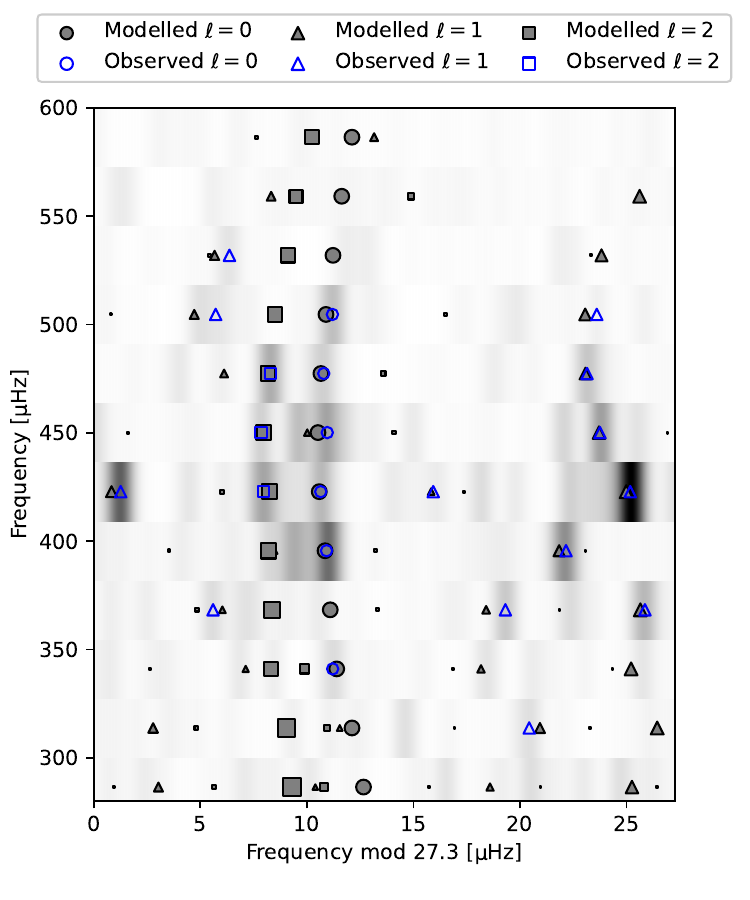}
\caption{Reconstructed power spectrum of \baql\ (greyscale) with the observed frequencies (open blue symbols; Table~\ref{tab:freqs}) and model frequencies (filled black symbols). For the latter, the symbol sizes for the non-radial modes are inversely proportional to mode inertia, scaled to the inertia of adjacent radial modes. }
\label{fig:model-echelle}
\end{figure}

\begin{table}
\caption{Properties of \baql.}
\label{tab:properties}
\centering
\begin{tabular}{lll}
\hline\hline
Property & Value & Source \\
\hline
\Teff         & $5092 \pm 50$\,K           & literature (Sec.~\ref{sec:properties}) \\
\mbox{[Fe/H]} & $-0.21\pm 0.07$\,dex               & literature (Sec.~\ref{sec:properties}) \\
radius        & $3.096 \pm 0.015$\,\Rsolar & literature (Sec.~\ref{sec:properties}) \\
mass          & $1.24\pm0.02$\,\Msolar            & this work (Sec.~\ref{sec:models}) \\ 
age           & $4.77\pm0.50$\,Gyr                & this work (Sec.~\ref{sec:models}) \\
$\log g$          & $3.549\pm0.002$\,            & this work (Sec.~\ref{sec:models}) \\ 
Density $\rho$           & $0.04166\pm0.00004$\,$\rho_\odot$               & this work (Sec.~\ref{sec:models}) \\
$\alpha_{\rm MLT}$           & $2.10\pm0.15$                & this work (Sec.~\ref{sec:models}) \\
$Y_{\rm init}$           & $0.24\pm0.02$                & this work (Sec.~\ref{sec:models}) \\
\hline
\end{tabular}
\end{table}

\subsection{Stellar parameters}
\label{sec:model-parameters}

\new{Figure~\ref{fig:hrd} shows the position of \baql\ in the Hertzsprung--Russell diagram, as well as the grid of theoretical models, confirming that the star is at the base of the red giant branch.}
The frequencies of the best-fitting model are shown in Fig.~\ref{fig:model-echelle}  (filled grey symbols), and we see an excellent match with most of the observed frequencies (open blue symbols). In Table~\ref{tab:properties} we list the stellar parameters, including mass, age, $\log g$, mean density, $\alpha_\text{MLT}$, and $Y_\text{init}$. We note that the uncertainties of $\log g$ and density are probably underestimated, since the systematic uncertainties can be many times larger than the statistical uncertainties reported here \citep[e.g.][]{Huber2022}. 

There has been considerable discussion about whether the choice of weighting terms appearing in Equation~\ref{eq:chi2} (such as $1/N_l$) can affect the derived stellar parameters \citep[e.g.][]{Cunha2021}. To investigate this, we tested an alternative approach where $1/N_l$ was replaced with a value of 1, effectively treating each mode frequency as equally important as each classical observable. This adjustment resulted in no statistically significant changes in the estimated mass and age, showing the robustness of these derived stellar parameters against the choice of weighting scheme.

We determined a mass of $1.24\pm0.02$\,\Msolar{}. Recent studies using only classical observables have reported less precise results. For example, \citet{Ghezzi2015} reported $1.140 \pm 0.105 $~\Msolar{} with PARSEC models \citep{Bressan2012}, and \citet{Karovicova++2022} obtained $1.36 \pm 0.13$~\Msolar{} with Dartmouth stellar evolution tracks \citep{Dotter2008}. 
These comparisons demonstrate the significant improvement in precision when asteroseismic data are incorporated. 
\new{However, we caution that the reported precision is not entirely realistic for several reasons: (i) the observational frequency uncertainties are often very small, which disproportionately dominates the likelihood function compared to the classical constraints \citep{Cunha2021}; (ii) the model uncertainties in the frequencies typically dominate, yet they are difficult to quantify when relying on a single set of models with limited variations in input physics \citep{SilvaAguirre2020, Christensen-Dalsgaard2020, Li-Yaguang++2024}; and (iii) the model-predicted frequencies are often correlated, but these correlations are rarely incorporated into the analysis \citep{Aerts++2008, Li-tanda2023}.}

The value of \numax\ implied by our model fits is $423\pm5 \,\muHz$, based on the standard scaling relation of $\numax \propto g/\sqrt{\Teff}$ \citep{Brown++1991, Kjeldsen+Bedding1995}, which falls within the range of measured values (see Fig.~\ref{fig:power-spectra}).
We also note that we used the measured interferometric radius of \baql\ as an input to the modelling.  If this measurement is not used, the stellar mass and radius agree with those listed in Table~\ref{tab:properties} but with uncertainties increased by about a factor of two.

Our asteroseismic analysis gives the age of \baql\ to be $4.77\pm0.50$\,Gyr, which again is consistent with classically determined stellar ages but more precise.  For example, an age of $5.86 \pm 1.87$~Gyr was reported by \citet{Ghezzi2015}. We note that our result also gives an accurate age for the M-dwarf companion, \baql~B (see Sec.~\ref{sec:properties}).

Including the interferometric radius in the fit has allowed us to constrain the mixing-length parameter quite well. At the base of the RGB, determining the radius with asteroseismology using individual frequencies alone (without $\nu_{\text{max}}$) is challenging due to the strong correlation between the radius and $\alpha_{\text{MLT}}$ \citep{Li-Yaguang++2024}. An accurate angular diameter makes \baql\ an important calibrator for $\alpha_{\text{MLT}}$ in this region of parameter space. We determined $\alpha_{\text{MLT}}$ for \baql\ to be $2.10 \pm 0.15$. The uncertainty is comparable to the grid step size, which suggests that the precision could be further improved by using finer sampling in the $\alpha_{\text{MLT}}$ parameter space. 
Hydrodynamical 3D simulations suggest that stars with properties similar to \baql{} should have values similar to that of the Sun \citep{Trampedach2014, Magic2015}. In fact, our value for \baql\ is slightly higher than our solar-calibrated value of $1.9$, and a similar difference between 1D modelling and 3D simulations has been reported from analysis of Kepler stars \citep{tayar2017, Viani2018, Li-Tanda++2018, Joyce2023}.

Models that excluded mixed-mode frequencies failed to constrain the initial helium abundance ($Y_{\text{init}}$). However, when mixed-mode information was included, $Y_{\text{init}}$ became better constrained. We found that using the g-mode period spacing alone was insufficient to achieve this constraint. Instead, the improvement is attributed to the g-mode phase ($\epsilon_g$), presumably through the modulation of the location of the g-mode cavity from chemical abundances. This aspect clearly warrants further detailed investigation.

\subsection{Comparison with M67}
\label{sec:m67}

\baql\ is in a similar evolutionary state to some members of the solar-metallicity open cluster M67, for which \citet{Reyes++2024} recently determined an age of $3.95^{+0.16}_{-0.15}$\,Gyr. 
Using photometry from \kepler/K2, \citet{Reyes++2025} have detected oscillations in more than 30 subgiants and RGB stars in M67, and we have compared them with our results for \baql, and a representative stellar track of a 1.22 M$_{\odot}$, [Fe/H]$=-0.21$ star, created and surface-corrected as described in that paper.

\begin{figure}
\centering
\includegraphics[width=\hsize]{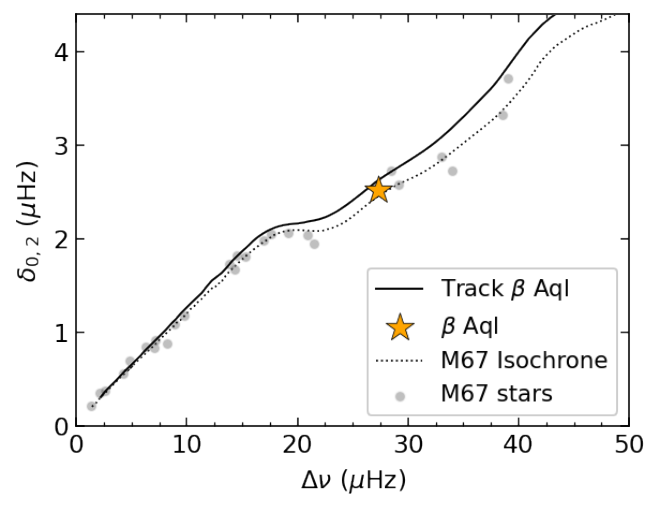}
\includegraphics[width=\hsize]{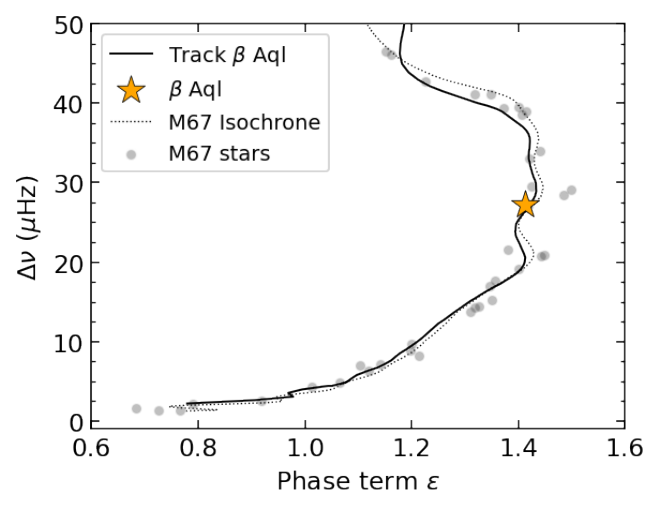}
\caption{Small frequency separations $\delta_{0,2}$ (top) and phase term $\epsilon$ (bottom) sequences from M67 stars, taken from \citealt{Reyes++2025-nature} and \citealt{Reyes++2025}, respectively. \baql\ is shown with a star symbol.}
\label{fig:M67}
\end{figure}

Figure~\ref{fig:M67} shows the comparison on two important asteroseismic diagrams. The upper panel shows the so-called C-D diagram \citep{Christensen-Dalsgaard1988}, which plots the small separation versus the large separation.  The small frequency separation, $\delta\nu_{02}$, is the mean difference between modes of degrees $\ell=0$ and~$2$, which we measured for \baql\ by vertically collapsing the \echelle\ diagram and fitting Lorentzian functions to the two ridges \citep[e.g.,][]{Bedding++2010}. \new{This
method boosts the signal-to-noise ratio of $\ell = 0$ and $\ell = 2$ modes while also reducing the impact of mixed modes on the final $\delta\nu_{02}$ measurements.} 
The lower panel plots \Dnu\ against the phase term,~$\epsilon$, which measures the absolute position of the mode pattern found from the above fit of the radial ridge \citep{Christensen-Dalsgaard1988, Huber++2010, White++2011-kepler, White++2011-diagrams, Kallinger++2012, Christensen-Dalsgaard++2014}\new{, as shown by the asymptotic relation for acoustic modes \citep{Tassoul1980}: 
\begin{equation}
    \nu _{n\ell} \simeq \Delta \nu \left(n + {\ell/2} + \epsilon \right) - \delta\nu_{0\ell},
\end{equation}
where $\nu _{n\ell}$ is the frequency of the mode of radial order $n$ and degree $\ell$}. 

The seismic properties of \baql\ align remarkably well with the sequence observed in M67, despite the difference in metallicity ([Fe/H] for M67 is close to zero). The wiggles seen in the C--D diagram depend on the location of the bottom of the convection zone, which in turn depends on mass and metallicity \citep{Reyes++2025-nature}. Because \baql\ is slightly lower in both metallicity and mass than the M67 giants, it still falls close to the M67 sequence in the diagram. The star's location in this particular region of the diagram would suggest it is well-suited to investigate the amount of overshooting at the bottom of the convective envelope \citep{Ong++2025, Reyes++2025-nature}. However, at the evolutionary stage of \baql\ we expect some scatter in the observed $\delta\nu_{02}$ values due to mixed modes. This results in \baql\ effectively blending in with the M67 sequence.
At the evolution stage, metallicity, and mass of \baql, the $\epsilon$ diagram is not expected to show clear sensitivity to metallicity and mass \citep{White++2011-kepler}, which is confirmed by the alignment with the M67 sequence. 

\section{Conclusions}

We have presented time-series radial velocities of the G8 subgiant star \baql\ obtained in 2022 and 2023 using SONG-Tenerife and SONG-Australia. We describe our method for processing the time series to  assign weights to the observations and remove bad data points. The resulting power spectrum clearly shows solar-like oscillations, and these are also seen in the power spectra of the 2022 TESS light curve and of short RV time series obtained with HARPS (in 2008) and SARG (in 2009). 

\baql\ is only the second star, after Procyon, for which we have observations of solar-like oscillations simultaneously in RV and photometry. We measured the ratio between the bolometric photometric amplitude and the velocity amplitude to be $26.6 \pm 3.1$\,ppm/\ms\ (Sec.~\ref{sec:phot-vel-ratio}), which compares with our measurement for the Sun (made using published data from SOHO/VIRGO and SONG-Tenerife) of $19.5 \pm 0.7$\,ppm/\ms\ (Sec.~\ref{sec:solar}).  Once the difference in effective temperatures of the two stars is taken into account, these measurements agree within their uncertainties.  We also measured the phase shift of the oscillations between SONG RVs and TESS photometry to be $-113^{\circ} \pm 7^{\circ}$, which agrees within uncertainties with the value in the Sun and with a 3-D simulation (Sec.~\ref{sec:overlap}).
Importantly, these results indicate that high-cadence photometric observations with TESS could be used to mitigate the effect of oscillations on RV exoplanets searches.

We extracted frequencies for \new{22} oscillation modes with angular degrees of $\ell=0$, 1 and~2 (Table~\ref{tab:freqs}). We carried out asteroseismic modelling by comparing the observed properties of \baql\ with a grid of models, yielding an excellent fit to the frequencies. The resulting stellar parameters are listed in Table~\ref{tab:properties}. In particular, we have been able to place quite strong constrains on the mixing length parameter ($\alpha_{\rm MLT}$) by including the interferometric radius in the model fitting.
We also found that the oscillation properties of \baql\ are very similar to subgiants and low-luminosity RGB stars in the open cluster M67 (Fig.~\ref{fig:M67}).
Further observations of \baql\ with SONG will be valuable, especially since the Mount Wilson S-index measurements show evidence for a magnetic activity cycle with a period of 4.7\,yr (Sec~\ref{sec:activity}).

\begin{acknowledgements}
The SONG network of telescopes is operated by Aarhus University, Instituto de Astrofísica de Canarias, the National Astronomical Observatories of China, University of Southern Queensland and New Mexico State University. 
We are grateful to the observing and technical support at each of the sites. We especially acknowledge Antonio Pimienta for his role as the curator of the Hertzsprung SONG Telescope at the Observatorio del Teide continuously for the past 15 years. Funding for the Stellar Astrophysics Centre was provided by The Danish National Research Foundation (Grant agreement no.: DNRF106).
We gratefully acknowledge support from the Australian Research Council through: LIEF Grant LE190100036; Discovery Projects DP220102254 (SLM) and DP210101299 (JL); Future Fellowships FT210100485 (SJM) and FT200100871 (DH); and Laureate Fellowship FL220100117 (TRB).
PLP acknowledges support from the Spanish Ministry of Science and Innovation with the grant no.\ PID2023-146453NB-100 (\textit{PLAtoSOnG}).

The HK\_Project\_v1995\_NSO data derive from the Mount Wilson Observatory HK Project, which was supported by both public and private funds through the Carnegie Observatories, the Mount Wilson Institute, and the Harvard-Smithsonian Center for Astrophysics starting in 1966 and continuing for over 36 years. These data are the result of the dedicated work of O. Wilson, A. Vaughan, G. Preston, D. Duncan, S. Baliunas, and many others.

SOHO is a project of international cooperation between ESA and NASA.

This work made use of several publicly available {\tt python} packages: {\tt astropy} \citep{astropy:2013,astropy:2018}, 
{\tt echelle} \citep{echelle2020},
{\tt lightkurve} \citep{lightkurve2018},
{\tt matplotlib} \citep{matplotlib2007}, 
{\tt numpy} \citep{numpy2020}, and 
{\tt scipy} \citep{scipy2020}.
\end{acknowledgements}

\ifarxiv
    
\else
    \bibliographystyle{aa}
    \bibliography{references}
\fi



\end{document}